\documentclass[print]{aa} 
\usepackage{savesym}
\usepackage{graphicx}
\pdfoutput=1
\usepackage{natbib}
\bibpunct{(}{)}{;}{a}{}{,} 
\usepackage{footnote}
\usepackage{lineno}
\usepackage{mathtools}
\usepackage{breqn}
\usepackage{amsmath}
\usepackage{xspace}
\usepackage{units}
\usepackage{relsize}
\usepackage{tikz}
\usepackage{amssymb,amsfonts,amsmath,amstext,amsgen,amsopn,amsxtra,indentfirst,times}
\usetikzlibrary{positioning}
\usetikzlibrary{shapes,arrows}
\usepackage{nicefrac}

\newcommand{\prior}{p({\bf m})\xspace}
\newcommand{\m}{{\bf m}\xspace}
\newcommand{\dat}{{\bf d}\xspace}
\newcommand{\post}{p({\bf m}|{\bf d})\xspace}
\newcommand{\fesi}{{\rm Fe}/{\rm Si}_{\bigcirc}\xspace}
\newcommand{\mgsi}{{\rm Mg}/{\rm Si}_{\bigcirc}\xspace}
\newcommand{\fesistar}{{\rm Fe}/{\rm Si}_{\mathlarger{\mathlarger \star}}\xspace}
\newcommand{\mgsistar}{{\rm Mg}/{\rm Si}_{\mathlarger{\mathlarger \star}}\xspace}
\newcommand{\fesima}{{\rm Fe}/{\rm Si}_{\mathlarger{\mathlarger{ \circledcirc}}}\xspace}
\newcommand{\mgsima}{{\rm Mg}/{\rm Si}_{\mathlarger{\mathlarger{ \circledcirc}}}\xspace}
\newcommand{\shann}{H_{\rm posterior}/H_{\rm prior}\xspace}
\newcommand{\si}{{\rm Si}_{\mathlarger{\mathlarger{\circledcirc}}}\xspace}

\begin{document}

   \title{Can we constrain interior structure of rocky exoplanets from mass and radius measurements?}
\titlerunning{Can we constrain interior structure of rocky exoplanets from mass and radius?}
\authorrunning{Dorn et al.}
   \author{Caroline Dorn\inst{1} 
   \and
   Amir Khan\inst{2}
       \and Kevin Heng\inst{1}
           \and Yann Alibert\inst{1}
              \and James A. D. Connolly\inst{3}
               \and Willy Benz\inst{1}
                       \and Paul Tackley\inst{2}
          }

   \institute{Physics Institute, University of Bern, Sidlerstrasse 5, CH-3012, Bern, Switzerland\\
              \email{caroline.dorn@space.unibe.ch}
         \and
             Institute of Geophysics, ETH Z\"urich, Sonneggstrasse 5, 8092 Z\"urich
             \and
             Institute of Geochemistry and Petrology, ETH Z\"urich, Clausiusstrasse 25, 8092 Z\"urich\\
             }

   \date{Received September 2014}

 
  \abstract
   {}
   {We present an inversion method based on Bayesian analysis to constrain the interior structure of terrestrial exoplanets, in the form of chemical composition of the mantle and core size. 
Specifically, we identify what parts of the interior structure of terrestrial exoplanets can be determined from observations of mass, radius, and stellar elemental abundances. }
   {We perform a full probabilistic inverse analysis to formally account for observational and model uncertainties and obtain confidence regions of interior structure models. This enables us to characterize how model variability depends on data and associated uncertainties.}
   {We test our method on terrestrial solar system planets and find that our model predictions are consistent with independent estimates. Furthermore, we apply our method to synthetic exoplanets up to 10 Earth masses and up to 1.7 Earth radii as well as to exoplanet Kepler-36b.
Importantly, the inversion strategy proposed here provides a framework for understanding the level of precision required to characterize the interior of exoplanets.}
   {Our main conclusions are: 
(1) observations of mass and radius are sufficient to constrain core size;
(2) stellar elemental abundances (Fe, Si, Mg) are key constraints to reduce degeneracy in  interior structure models and to constrain mantle composition;
(3) the inherent degeneracy in determining interior structure from mass and radius observations does not only depend on measurement accuracies but also on the actual size and density of the exoplanet.
We argue that precise observations of stellar elemental abundances are central in order to place constraints on planetary bulk composition and to reduce model degeneracy.
 We provide a general methodology of analyzing interior structures of exoplanets that may help to understand how interior models are distributed among star systems.
The methodology we propose is sufficiently general to allow its future extension to more complex internal structures including hydrogen- and water-rich exoplanets.}


   \maketitle

\section{Introduction}

\sloppy
Major advances in detection and characterization of exoplanets have been achieved over the past decade. To date, several hundred
have been characterized in terms of mass and radius by space-based or ground-based observations. 
Continued improvements in observational techniques allow for more precise inference of mass and radius. To date, even a few small-mass exoplanets ($<$ 10 Earth masses ($M_E$)) with uncertainties in mass and radius  below 20\% and 10\%, respectively, have been discovered (e.g., Kepler-36b, CoRot-7b) \citep[e.g.,][]{pepe}.
Space-based missions, like {\it Kepler} \citep{borucki} and {\it CoRoT} \citep{borde}, have been able to measure the transits of these objects and infer the radius. Within the first 16 months of the {\it Kepler} mission alone, 207 planetary candidates of radius $R < 1.25$ Earth radii ($R_E$) and 680 super-Earth-sized ($1.25R_E < R < 2R_E$) planetary candidates were reported by \cite{batalha}.

From knowledge of mass and radius we are able to derive constraints on the interior structure of exoplanets. Previous studies concerned with rocky exoplanets \citep[e.g.,][]{valencia06, sotin07, seager2007, fortney, wagner11,Mocquet}  have generally concentrated on computing mass-radius relations based on terrestrial-type interior structures and compositions. These studies showed that different interior models are capable of explaining the observations within their uncertainties. Such ``forward" approaches, however, do not quantify the inherent degeneracy of interior structure models. For example, \citet[][]{valencia06, valencia07b, sotin07, wagner11} it has been shown  that different core sizes and mantle compositions affect the mass-radius relationship, but to our knowledge no comprehensive study of the degeneracy of these model parameters has been performed.  
As a consequence, it is not fully understood which interior-structure parameters can be estimated from the data and which structure parameters tend to be strongly correlated. \\
\indent
In the light of the inherent ambiguity in determining interior structure from mass and radius \citep[e.g.,][]{howe,rogers2010}, analysis is most sensibly conducted in an inverse sense \citep[][]{Mo2002}. We therefore propose a complete Bayesian inverse analysis by employing a Markov chain Monte Carlo (McMC) method to provide full probability distributions for the model parameters of interest (mantle composition and core size).
%


Early studies of mass-radius relations \citep{zapolsky} considered only homogeneous and simple monoatomic bulk compositions (H, He, C, Fe, Mg, and H/He mixtures). More recent studies assume differentiated terrestrial-like compositions \citep[e.g.,][]{valencia06, sotin07, seager2007, fortney}. Based on 
solar system observations and planet formation and differentiation models, terrestrial planets are generally thought to be differentiated in an iron-rich core, a silicate mantle and a crust \citep[see][for a review of previous and current work]{howe}. In this work, we focus on terrestrial-type rocky planets and assume that these consist of a pure iron core, a silicate mantle comprising the oxides Na$_2$O-CaO-FeO-MgO-Al$_2$O$_3$-SiO$_2$, and that volatiles have a negligible effect on mass and radius.

In order to be able to compute the radius for a planet of a given mass and composition (the ''forward" model) we solve the hydrostatic equilibrium equation coupled with a thermodynamic approach based on Gibbs free-energy minimization and Equation-of-State (EoS) modeling. In this regard, our work is similar to earlier studies \citep{valencia07a, sotin07, seager2007, fortney}, except for the use of thermodynamic modeling. This addition allows us to compute from first principles of thermodynamics,  the density profile of the planet for an arbitrary bulk silicate mantle composition. Moreover, given that cosmochemically-derived elemental abundances (e.g., Fe, Mg, and Si) have proved important as constraints on the composition of the terrestrial planets  \citep[e.g.,][]{grasset09, rogers2010}, we investigate the effect of including abundance constraints in determining interior structure.

Although similarities between the present work and the Bayesian analysis of \cite{rogers2010} exist, our approach nonetheless differs in several respects: 
\begin{itemize}  \itemsep0pt
\item our inversion method is applicable to planets of any mantle composition and core structure;

\item our inversion method is general in that it can be easily adapted to more complex planetary models that include oceans and atmospheres;

\item  we infer probability distributions on interior-structure parameters (core radius and mantle composition) 
for a wide range of masses ($0.1 M_E<M<10 M_E$) and radii ($0.5 R_E<R<1.7 R_E$);

\item  we demonstrate how observations of a few key elemental abundances (e.g., Fe, Mg, Si) of the host star's photosphere significantly reduces the degeneracy of interior structure models;

\item we quantify the influence of measurement uncertainties on the determination of interior 
structure models.

\end{itemize}


The outline of this study is as follows: In Section \ref{Methodology} we introduce the physical model that links data and model parameters, and outline the inversion strategy. In Section \ref{Vali}, we first test our method on Earth using Earth bulk elemental abundances derived from geochemical analysis of mantle rocks. Next, we apply the method to all terrestrial solar system planets. In Section \ref{Results}, we present results for synthetic exoplanets and Kepler-36b followed by discussion and conclusions.

\section{Methodology}
\label{Methodology}

\subsection{Abundance constraints}
 
 In summary, the importance of additional constraints in the form of chemical abundances of the elements Fe, Mg, Si are investigated in detail. 
 Arguments from planet formation favor certain interior compositions as discussed by e.g., \citet[][]{grasset09, rogers2010}. 
 
   Observations and theoretical considerations suggest that exoplanet bulk composition is dictated by stellar composition: First, relative elemental abundances of Fe, Mg, and Si are similar among the Sun, Earth, Mars, the Moon as well as meteorites \citep{lodders03,drake,mcdono}. Meteorites are believed to be chemically similar to the building blocks of planets \citep{morgan}, both being condensates from the solar nebula that experienced the same fractionation processes. 
Second, it has been demonstrated that during planet formation, planetary bulk and stellar ratios of Fe/Si and Mg/Si are very similar \citep[e.g.,][]{bond,elser,johnson,thiabaud} since the refractory elements Fe, Si, and Mg condense at similar temperatures corresponding to small distances to the host star ($\sim$ 1 AU).

For the Sun and our solar system planets, bulk Fe/Si and Mg/Si (henceforth $\fesi$ and $\mgsi$) have been inferred from photospheric analysis and from geochemical analysis of meteoritic and cosmochemical material, respectively.
For extrasolar systems, we shall assume that $\fesi$ and $\mgsi$ equal to their stellar counterparts $\fesistar$ and $\mgsistar$.

\begin{figure*}
\centering
 \includegraphics[width = .3\textwidth, trim = 0cm 11cm 15cm 0cm, clip]{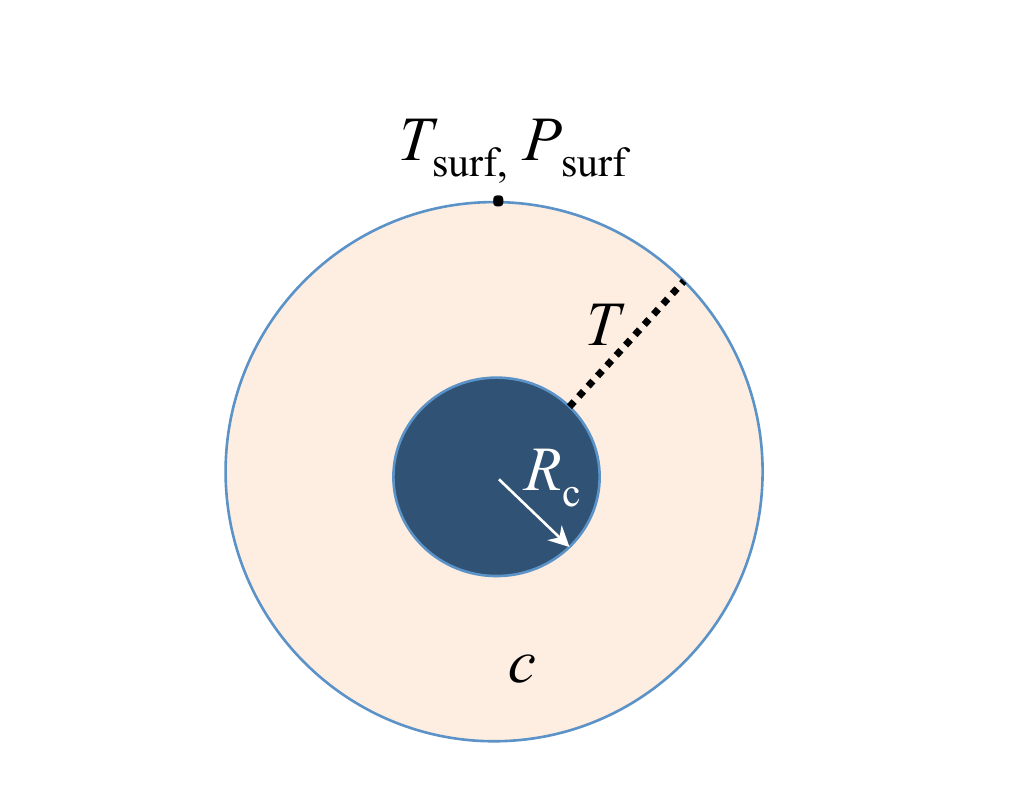}
 \caption{Illustration of model parameterization. Parameters are core radius $R_c$ and mantle composition $c$ comprising the oxides Na$_2$O-CaO-FeO-MgO-Al$_2$O$_3$-SiO$_2$. Mantle temperature profile $T$, surface temperature $T_{\mathrm surf}$ and pressure $P_{\mathrm surf}$ are fixed input parameters. In the iron core an adiabatic temperature profile is used.  \label{illustration}}

\end{figure*}

\begin{figure*}
\center
 \includegraphics[width = 1\textwidth, trim = 1cm 17.5cm 7cm 2cm, clip]{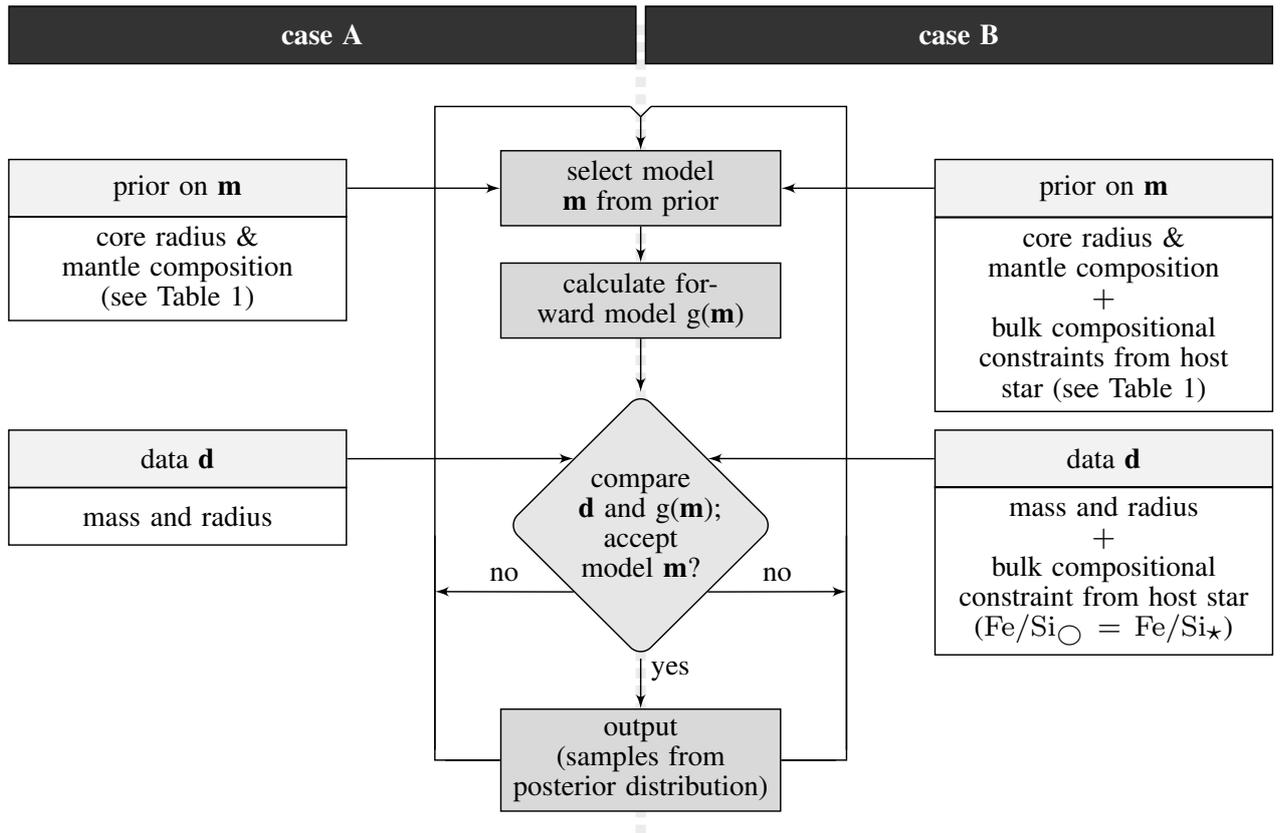}
 \caption{Inversion scheme. Note that the bulk compositional constraints from the host star (case B) used both in prior and data are independent. The iterative scheme is repeated until the output converges to the posterior distribution. See section \ref{Methodology} for more details.  \label{flowchart1}}
\end{figure*}

\subsection{Data}
\label{data}

The aim of our study is to characterize the interior structure of terrestrial exoplanets based on inferred mass and radius from radial velocity and transit observations and constraints on bulk Fe, Mg, and Si abundances ($\fesi$ and $\mgsi$) inferred from stellar photospheric analysis. Mass, radius and bulk abundance constraints constitute our data \dat. 
For the purpose of illustrating the relative importance of data on inverted model parameters, we consider the following two cases: 
\begin{description}
\item[\bf Case A] Data are mass $M$ and radius $R$. 
\item[\bf Case B] Data are mass $M$, radius $R$, and compositional constraints ($\fesi$ and $\mgsi$) obtained from stellar spectroscopic observations of the host star. 
\end{description}

Note that $\fesi$ expresses the mass ratio between the mass of iron to silicate for the entire planet (core and mantle). Given that we assumed a pure iron core, this ratio is equal to the mass of iron in core and mantle divided by the mass of silicate in the mantle $M_{\rm si, mantle}$. This can be written as follows:

\begin{equation}
\fesi = \fesima + M_{\rm core}/M_{\rm Si, mantle}
\label{fesi}
\end{equation} 
where $M_{\rm core}$ is mass of iron in the core, and $\fesima$ denotes the mass ratio of Fe/Si in the mantle. Mantle Mg/Si is referred to as $\mgsima$. Since all Si and Mg is assumed to be in the mantle, $\mgsi = \mgsima$.

\subsection{Model parameterization}
\label{model}

Our exoplanet interior model consists of a layered sphere with an iron core surrounded by a silicate mantle (see Figure \ref{illustration}). Model parameters include core size $R_{\rm c}$ and mantle silicate composition using the NCFMAS model chemical system that comprises the oxides Na$_2$O-CaO-FeO-MgO-Al$_2$O$_3$-SiO$_2$ and accounts for more than 98\% of the mass of Earth’s mantle \citep{Irifune} (see Appendix \ref{fmssystem}). We parameterize mantle composition by mantle Si-content defined as $M_{\rm Si, mantle}/M_{\rm mantle}$, (referred to as $\si$), $\fesima$ and $\mgsima$.  Na$_2$O, CaO, and Al$_2$O$_3$ are minor components and their relative abundance can be constrained from stellar photospheric observations \citep[e.g.,][]{elser}.  In this study, we assume chondritic abundances (see Table \ref{table_range}). The set of parameters that constitutes our model parameter \m is thus:
\begin{itemize}
\item core radius $R_c$,
\item mantle Si-content $\si$,
\item mantle $\fesima$,
\item mantle $\mgsima$.
\end{itemize}
The mantle composition parameters $\si$, $\fesima$, and $\mgsima$ depend on each other, i.e. the fractions of oxides in the NCFMAS system must sum up to one. All parameters are assumed to be uniformly distributed unless stated otherwise (see Table \ref{table_range}). 

\subsection{Prior information}
\label{prior}

\begin{savenotes}
\begin{center}
\begin{table*}[ht]
\caption{Prior ranges of model parameters for case A \citep{Beirao, gilli, thiabaud} and case B \citep{lodders03}. Uniform distributions unless stated otherwise. Si, Fe, and Mg are in wt\% (weight percentage). Minor components of Na$_2$O, CaO, and Al$_2$O$_3$ are fixed to chondritic abundances of 2 wt\%, 2 wt\% and 4.8 wt\%, respectively \citep{lodders03}. \label{table_range}}
\begin{center}
\begin{tabular}{lccc}
\hline\noalign{\smallskip}
model parameter of $\m$ & case A & case B \\
\noalign{\smallskip}
\hline\noalign{\smallskip}
$\si$  & 8 --34 (non-uniform\footnotemark[5]) & 14 --30 (non-uniform\footnotemark[5]) \\
$\fesima$           & 0 --4.15 & 0 --2.23 \\
$\mgsima$          & 0.61--2.32 & 0.89 $\pm$ 0.08 (Gaussian)\\
core radius $R_c$        & 0-$R$	 & 0-$R$ \\
\hline
\end{tabular} \\
\footnotemark[5] see appendix \ref{priorsi},
\end{center}
\end{table*}
\end{center}
\end{savenotes}

For case A, prior bounds  have been chosen based on: (1) the assumption that the bulk metallic  composition of a planet is similar to the abundance of the host star \citep{thiabaud} and (2) the fact that chemical abundances of observed stars with planetary companions seem to fall within certain ranges \citep[e.g.,][]{Beirao, gilli}. Thus, the variability of observed elemental abundances in stars defines the prior bounds on $\fesima$ and $\mgsima$.

For case B, prior bounds on $\fesima$ and $\mgsima$ are derived from $\fesi$ and $\mgsi$ inferred from the host star's photospheric abundance: 
(1) Since all Si and Mg is assumed to be in the mantle, $\mgsi$ defines the prior bounds on $\mgsima$; 
(2) Fe, on the other hand, is distributed between core and mantle. Thus, the bulk constraint $\fesi$ defines only the upper bound of the prior on $\fesima$. For exoplanets, $\fesi$ and $\mgsi$ are defined by the host star's photospheric elemental abundance since we assume that $\fesi = \fesistar$ and $\mgsi = \mgsistar$. 
$\mgsistar$ is Gaussian-distributed.
We consider solar estimates for Fe, Si, and Mg \citep{lodders03} and associated uncertainties as representative of $\fesi$ and $\mgsi$ throughout this study unless stated otherwise.

In both cases, the prior bounds on $\si$ depend on the prior bounds of $\fesima$ and $\mgsima$, since the fractions of oxides in the NCFMAS chemical system must sum to one. This implies a non-uniform prior of $\si$ (see Appendix \ref{priorsi}).

Additional input parameters that are fixed in this study are (1) surface pressure and temperature, (2) thermal gradient in the mantle based on adiabatic gradient of Earth's mantle, and (3) absolute amount of minor components (Na$_2$O, CaO, and Al$_2$O$_3$) in the mantle.

Using a fixed temperature profile for the mantle is demonstrably not a limitation of our model. Many authors have shown that the variations due to temperature for rocky planets are small compared to uncertainties of material properties of silicate or iron compositions \citep[e.g.,][]{sotin07, seager2007, grasset09, valencia10}. This is demonstrated in Appendix \ref{tempvary} where it is shown that varying mantle temperatures introduces little variation in computed densities. As a consequence, we fix temperature profiles by using Earth's surface temperature, lithosphere  and mantle temperature gradient. \\


\subsection{Interior structure model}
\label{forward}

Data and model parameters are linked by a physical model embodied by the forward operator $g(\cdot)$

\begin{equation}
\dat = g(\m)
\label{equ_forward}
\end{equation}

For a given model $\m$, interior structure (density profile), the total mass, and $\fesi$ are computed for the purpose of comparing with observed data $\dat$.
Equation \ref{equ_forward} represents the forward problem, which is computed using thermodynamic method and Equation-of-State (EoS) modeling.  The complete solution of the forward problem is summarized as follows:

\begin{equation}
\{c, R_{\rm c}, T\} \stackrel{g_1}{\rightarrow} \mathcal{M} \stackrel{g_2}{\rightarrow}\rho \stackrel{g_3}{\rightarrow}\{M,R\} 
\end{equation}
where $c$ denotes the NCFMAS mantle composition, $T$ temperature, $\mathcal{M} $ equilibrium mantle mineralogy, and $\rho$ density (all parameters are a functions of radius). The forward operator $g_1$ embodies Gibbs free-energy minimization that computes stable mineral modes as a function of $P$ (pressure), $T$, and composition, $g_2$ calculates density, $g_3$ integrates $\rho$ over $R$ to compute $M$. We assume subsolidus conditions throughout.

\subsubsection{Thermodynamic modeling}
\label{thermodyn}
Possible mantle compositions are explored within the Na$_2$O-CaO-FeO-MgO-Al$_2$O$_3$-SiO$_2$ model system. To compute the mantle density profile, we employ a self-consistent thermodynamic method. For this purpose we assume: thermodynamic equilibrium; mantle mineral phases of large exoplanets are those that potentially occur in the deep mantle of the Earth; and, when required, the thermodynamic properties of the mineralogy can be extrapolated to more extreme conditions than realized on Earth. The equilibrium mineralogy and, consequently its density, is computed as a function of pressure, temperature, and bulk composition by Gibbs energy minimization \citep{connolly09}. For these calculations the pressure is obtained by integrating the load from the surface boundary and temperature is obtained by integrating an Earth-like temperature gradient from an arbitrarily assumed surface temperature. Mineral equations of state and parameters are as given by \citet{stixrudea,stixrudeb}. Because the equilibrium assumption is dubious at low temperature \citep[e.g.,][]{wood}, for models that require rock properties at temperatures below 800 K, the stable mineralogy is first calculated at 800 K and its physical properties are then computed at the temperature of interest.

\subsubsection{EoS for iron core}
\label{eos}
In order to compute the core density profile, we use the EoS for pure iron derived by \citet{belonoshko10}, which is similar to the Mie-Gr\"uneisen-Debye EoS \citep{jackson}, except for a different thermal pressure term. This EoS for iron is preferred, as it allows us to closely reproduce high-precision volumetric experiments of iron under high-pressure and high-temperature conditions.

The adiabatic gradient in the core is computed as 

\begin{equation}
{d T \over d P} ={ \gamma T \over K_S } 
\end{equation}
where $K_S$ is adiabatic bulk modulus, which is related to the isothermal bulk modulus $K_T$ through
$K_S = \left( 1 + \gamma \alpha T \right) K_T$. The 
parameters of the Mie-Gr\"uneisen-Debye equation
{$\gamma = \gamma_0 \left( \rho \over \rho_0 \right)^{-q}$}
 for iron are $\rho_0 = 8.334$ g/cm$^3$, $q = 0.489$, and $\gamma_0 = 2.434$. Density $\rho$ and thermal expansion coefficient $\alpha$ are then directly computed from Belonoshko's EoS.

\subsection{Inversion method}
\label{inversion}

Probabilistic inversion methods are suitable when data are sparse, the physical model ({Eq.\ \ref{equ_forward}}) is highly non-linear and/or it is expected that very different models are consistent with information and data \citep[e.g.,][]{MooT}. 
While stochastic sampling-based approaches for high-dimensional problems are computationally expensive, numerous global search methods exist. These include Markov chain Monte Carlo methods (McMC), simulated annealing or genetic algorithms. 
McMC has the advantage that parameter uncertainties are formally assessed. Furthermore, the McMC method generates samples whose distribution converges to a stationary distribution coinciding with the posterior probability density function (pdf) \citep[e.g.,][]{hastings}. Although the dimensionality of the presented inverse problem is low (4 parameters), offering the possibility of performing grid-search, we nonetheless employ McMC, because we intend to expand the complexity of the problem to higher dimensions, in the future.
For consistency, we have also used the grid-search method to verify our results.

\subsubsection{Bayesian analysis}
We employ a Bayesian method to compute the posterior probability density function (pdf) for each model parameter ($R_c$, $\si$, $\fesima$, $\mgsima$) from data ($M$, $R$ and $\fesi$) and prior information. According to Bayes' theorem, the posterior distribution for a fixed model parameterization \m, conditional on data \dat, is given by \citep{tarantola82}:

\begin{equation}
\label{post}
\post \propto \prior L({\bf m}|{\bf d}),
\end{equation}

where $p({\bf m})$ represents prior information on model parameter \m.  $L({\bf m}|{\bf d})$ is the likelihood function and can be interpreted in probabilistic terms as a measure of how well a model \m fits data \dat.

Assuming uncorrelated and normally-distributed residuals (see Eq. \ref{mis}), the likelihood defines a ``measure'' between \dat and \m and can be written as:

\begin{equation}
\label{like}
L({\bf m}|{\bf d}) = \frac{1}{(2\pi)^{N/2} (\prod_{i=1}^{N}\sigma_i^2)^{1/2}} {\rm exp}\left(- \frac{1}{2} \sum_{i=1}^{N}\frac{Q_i}{\sigma_i^2}\right)
\end{equation}
where $N$ is total number of data points, $\sigma_i$ is the estimated error of the i{\it th} datum, and $Q_i$
is data misfit, or squared residual, given by

\begin{equation}
\label{mis}
Q_i = (g_i({\bf m})-{\bf d_{\rm i}})^T(g_i({\bf m})-{\bf d_{\rm i}})
\end{equation}
where $g(\m)$ is the forward model discussed in section \ref{forward}.


\subsubsection{Markov chain Monte Carlo (McMC)}

For high-dimensional and non-linear inverse problems or sparse data, it is in practice
impossible to derive the posterior distribution analytically. 
McMC techniques offer an efficient method of performing Bayesian analysis (Eq. \ref{post}). McMC sampling methods iteratively
search the space of feasible solutions. The iteration scheme used is based on the Metropolis algorithm, which proceeds as follows:

\begin{enumerate}
\item An initial starting model $\m_{\rm old}$ is drawn at random by sampling from the prior distribution.
\item The posterior density of $\m_{\rm old}$ is calculated by evaluating the product of the likelihood of the corresponding forward model and prior density.
\item The current model parameter is perturbed to obtain $\m_{\rm new}$, which is subsequently created from a proposal distribution. Here, the proposal distribution is uniformly bounded and centered around $\m_{\rm old}$. Generally, it is chosen such that the size of the model perturbation allows for a reasonable rate of accepted transitions in the McMC procedure, typically around 30\% \citep{gilks}.
\item The proposal $\m_{\rm new}$ is accepted with probability \citep{MooT}: \\ $P_{accept} = min\{1, {\rm exp}(l(\m_{\rm new}|\dat) - l(\m_{\rm old}|\dat) )\}$
\item If the proposal is accepted the Markov chain moves to $\m_{\rm new}$, otherwise the chain remains at $\m_{\rm old}$.
\item Steps 2 to 5 are repeated.
\end{enumerate}

After many iterations, the accepted samples that are generated with this approach are (1) independent of
the starting model and (2) distributed
according to the posterior distribution. The efficiency of
sampling is strongly dependent upon the proposal distribution. If this distribution is badly chosen,
the acceptance rate of solutions might be unacceptably low,
resulting in very poor efficiency regarding convergence \citep{hastings}. On the contrary, if the proposal
distribution is well chosen, the McMC sampler will rapidly
explore the posterior target distribution.
Here, the posterior information is gathered from a large number of sampled models ($\sim 10^4$). Only the statistical nature of sampled model
features are of interest. Presently, we will concentrate on computing marginal posterior distributions in order to provide the reader with a notion of model parameter uncertainties. Information on single parameters  are obtained by one-dimensional (1D) marginals; higher dimensional marginal pdfs reveal the correlation that exists among several
parameters. If the data are able to constrain the model (i.e. informative), differences between prior and posterior pdfs are expected.

Given the large number of models that have to be computed, the calculation of the forward model must be computationally very efficient. In our work, generating the planet's internal structure takes on average 0.8 seconds of CPU time on a four quad-core AMD Opteron 8380 CPU node with 32 GB of RAM. In all, we sampled about $10^7$
models and retained around $10^4$ models for further analysis.

\section{Method validation}
\label{Vali}

\subsection{Synthetic case}

\begin{figure*}[ht]
\center
 \includegraphics[width = 1\textwidth, trim = 5cm 10.3cm 13cm 0cm, clip]{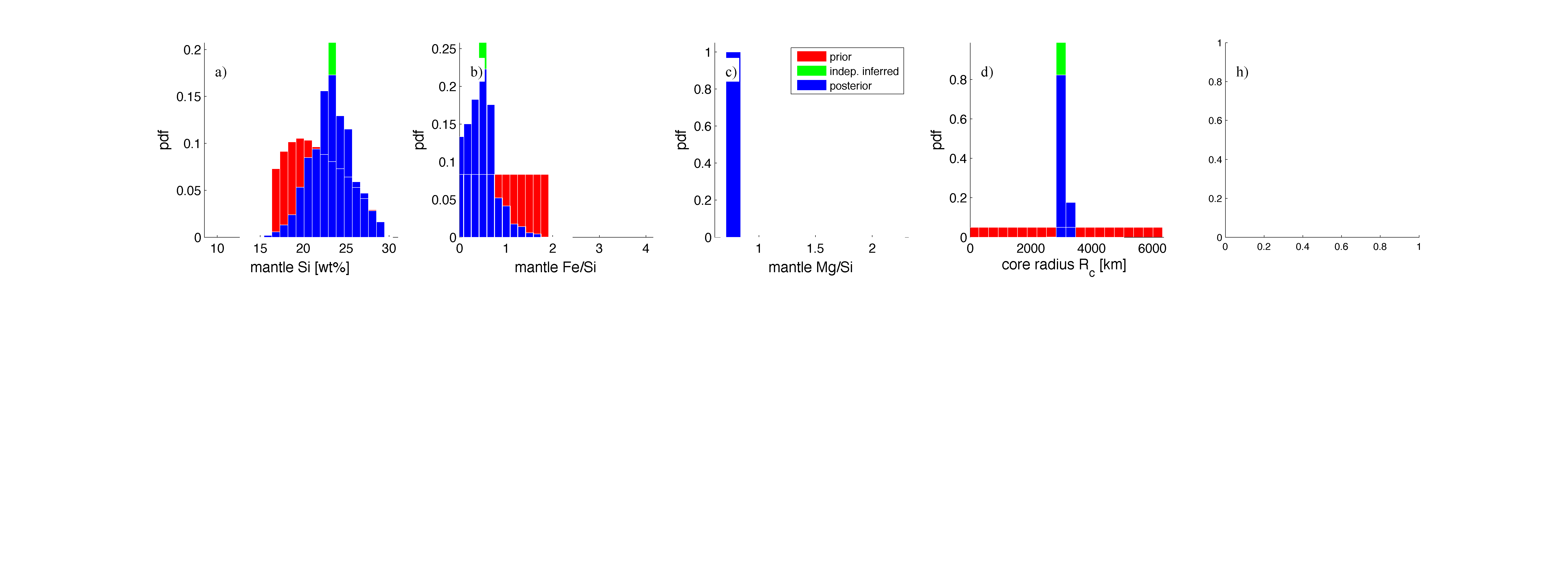}\\
 \caption{Sampled 1D marginal posterior distributions for the synthetic test where model parameters are inverted for an Earth-sized planet: prior (red), posterior (blue) and independent estimates (green) of model parameters: (a) mantle Si-content $\si$, (b) mantle $\fesima$, (c) mantle $\mgsima$, and (d) core radius $R_c$. \label{figure_synth}}
\end{figure*}

We apply our method to a synthetic planet that ultimately serves as a ''benchmark" test. Structure and composition of the synthetic model are defined as:
\begin{itemize}
\item $\si = 23.4$ wt\%
\item $\fesima = 0.5$; $\fesi = 1.69$
\item $\mgsima = \mgsi = 0.89$
\item $R_c  = 0.47*R$
\end{itemize}
Using the interior structure model (Section \ref{forward}) we calculate the synthetic data to be $M = 1 M_E$ and $R = 1 R_E$. For the inversion we consider the following data uncertainties:
$\sigma_{R} = 0$, 
$\sigma_{M} = 0.001 M$, together with $\fesistar = 1.69 \pm 0.18$ and $\fesistar = 0.89 \pm 0.08$.
Results are shown in Figure \ref{figure_synth} and demonstrate the ability of the method to retrieve the internal structure of the synthetic planet.
As expected, the most likely posterior model overlaps with the true model. The inherent degeneracy in interior structure is apparent from the spread of the posterior.

\subsection{Solar System planets}

We test our method on the terrestrial solar system planets Earth, Moon, Mars, Venus, and Mercury, for which a series of independent model parameter estimates are available from a host of geophysical, geo-, and cosmochemical data gathered in situ as well as from orbiting spacecraft (see Table \ref{table_true}). These tests are important inasmuch as they allow us to benchmark our method. 
For simplicity, Table \ref{table_true} only summarizes a single best estimate of independent interior models. Associated uncertainties are very different depending on data and model assumptions. For example, core radius $R_{\rm c}$ of the Moon, Mars and Mercury, is estimated to be $310--320$ km \citep{kuskov}, $1680 \pm 150$ km \citep{khan08}, and $2020 \pm 100$ \citep{padovan}, respectively.

In the following, we extensively test different scenarios for the Earth: case A and B. Input for cases A and B is defined in section \ref{Methodology}. 
These cases show that the compositional constraints $\fesi$ and $\mgsi$ are crucial for determining interior structure and that $\fesistar$ and $\mgsistar$ are good proxies for them. How well data for case B allow us to recover the interior structure is demonstrated for the Moon, Venus, Mars, and Mercury.
For all terrestrial planets we use $1\%$ and $0\%$ uncertainties on observed $M$ and $R$, respectively, and $\fesistar = 1.69 \pm 0.18$ \citep{lodders03}. Minor components of Na$_2$O, CaO, and Al$_2$O$_3$ are fixed to chondritic values \citep{lodders03}. Prior ranges for the various parameters are listed in Table \ref{table_range}.

\begin{center}
\begin{table*}[ht]
\caption{Independent best estimates of model parameters for terrestrial solar system planets. The preliminary reference earth model (PREM) refers to \citet{dziew}. \label{table_true}}
\begin{center}
\begin{tabular}{lccccc}
\hline\noalign{\smallskip}
planet &  $\si$  [wt\%]	& $\fesima$ 	& $\mgsima$ & $R_{\rm c}$ [km] & reference \\
\noalign{\smallskip}
\hline\noalign{\smallskip}
Earth & 22 		& 0.17		& 0.83 	& 3400		& PREM \\
Moon & 23 		& 0.4-0.47		& 0.8 		& 310-320		& \citet{kuskov} \\
Mars  & 21 		& 0.64		& 0.97 		& 1680		& \citet{khan08} \\
Venus& -- 		& --		& -- 		& 3100 		&  \citet{aitta} \\
Mercury& -- 		& --		& -- 		& 2020  	&  \citet{padovan, hauck} \\
\hline
\end{tabular}
\end{center}
\end{table*}
\end{center}

\paragraph{Earth}

\begin{figure*}[ht]
\center
 \includegraphics[width = 1\textwidth, trim = 5cm 0cm 13cm 0cm, clip]{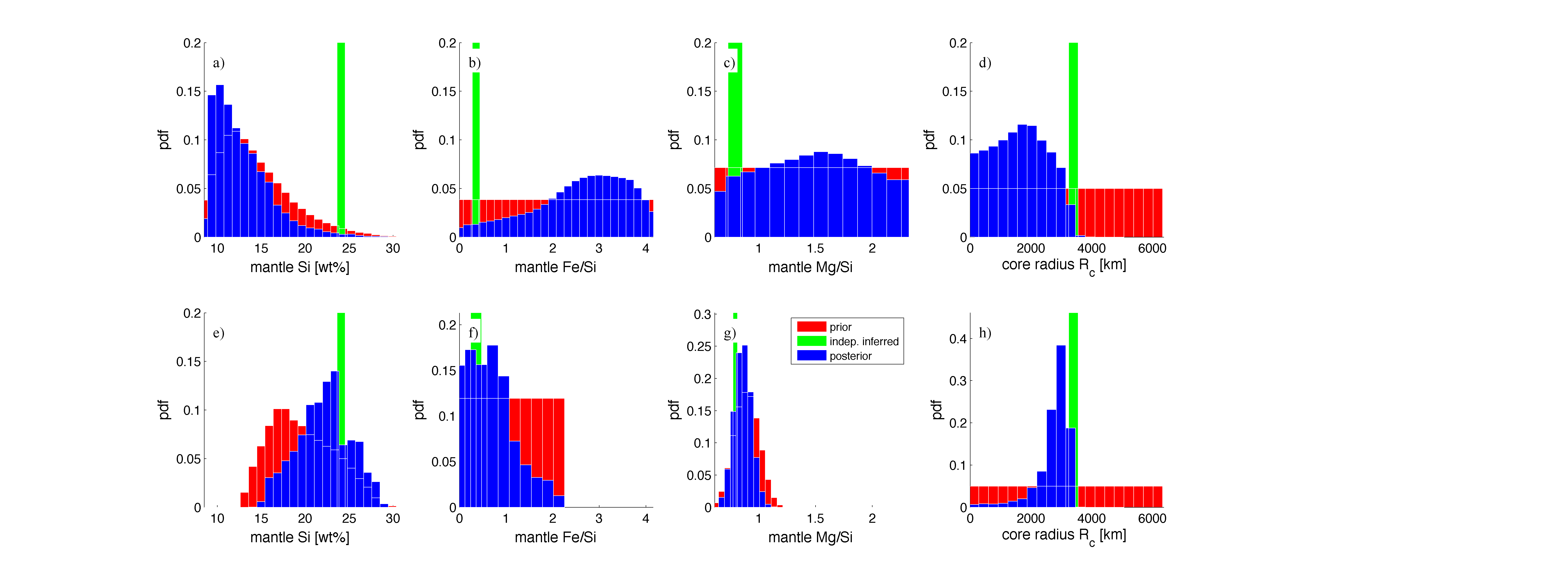}\\
 \caption{Sampled 1D marginal posterior distributions for Earth, (a-d) case A, (e-h) case B: prior (red), posterior (blue) and independent estimates (green) of model parameters: (a,e) mantle Si-content $\si$, (b,f) mantle $\fesima$, (c,g) mantle $\mgsima$, and (d,h) core radius $R_c$. Independent estimates are listed in table \ref{table_true}. Note that prior ranges are different for the two cases, but axis ranges are kept the same to ease comparison. \label{figure_earth}}
\end{figure*}

Results for the Earth for cases A and B are shown in Figure \ref{figure_earth}. 
Independent estimates of mantle composition and core size (Table \ref{table_true}) are shown as green bars.
Inversion of mass and radius only (case A) provides little information on mantle composition, but some constraints on $R_{\rm c}$.
 In contrast, addition of $\fesi$ and $\mgsi$ (case B) enables us, as expected, to constrain mantle composition and thereby obtain improved estimates of core size. Moreover, better agreement with the independent estimates is also observed.

  There is an underestimation of the core radius due to our assumed core composition.  In the PREM model \citep[][]{dziew}, the core is less dense which reflects the presence of lighter elements (e.g., S, Si, C, O) in addition to Fe-Ni.  Since we consider pure iron, we systematically overestimate core density and therefore underestimate core size. For both cases A and B, core sizes are $0 \le R_{\rm c} \le 2930$ km and $1470 \le R_{\rm c} \le 3300$ km (95\% credible intervals), respectively. 
Furthermore, we observe that $\mgsima$ is not well constrained by data.

\begin{figure*}[ht]
\center

 \includegraphics[width = .8\textwidth, trim = 2cm 1.3cm 0cm 1cm, clip]{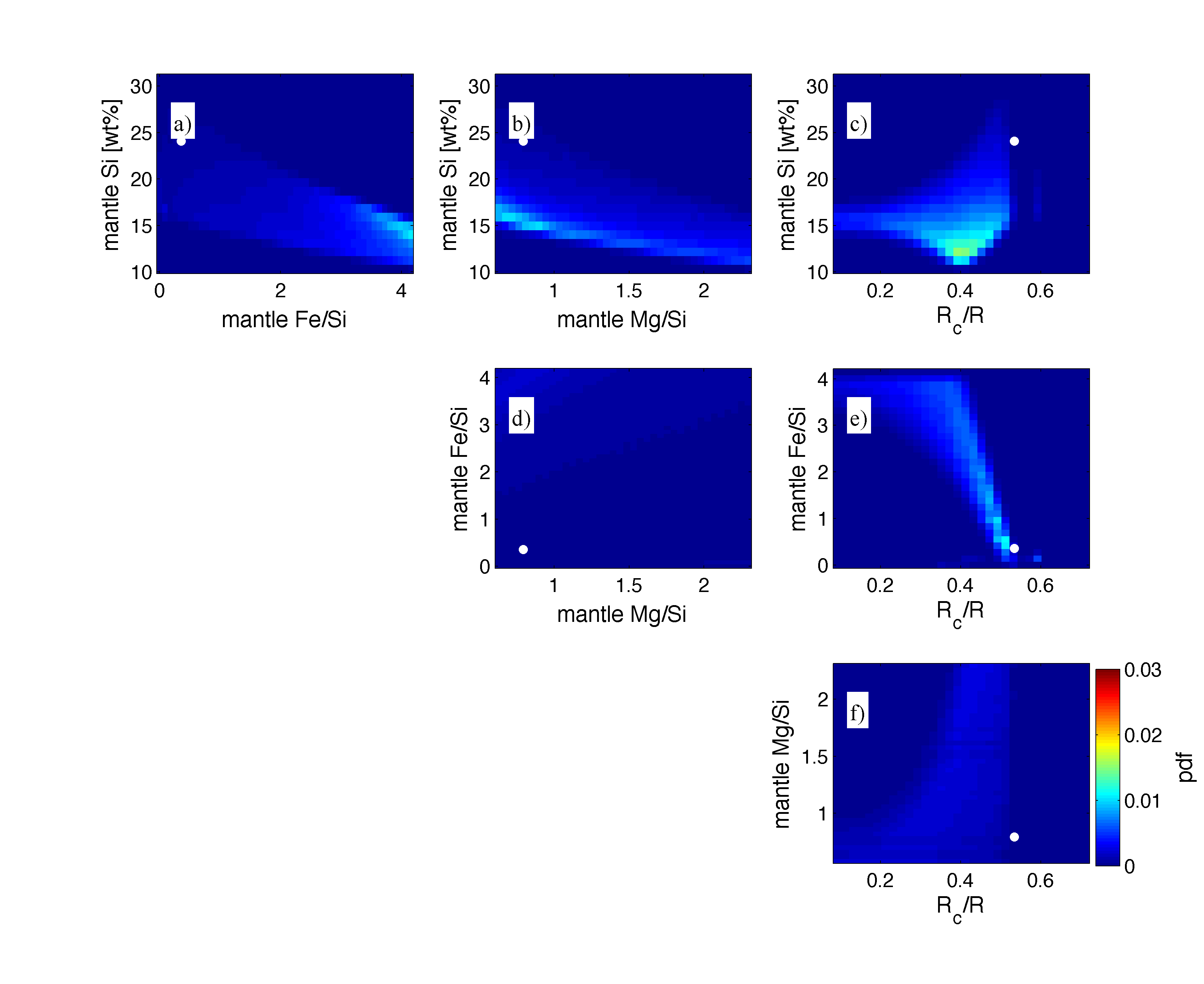}\\
 \includegraphics[width = .8\textwidth, trim = 2cm 1.3cm 0cm 1cm, clip]{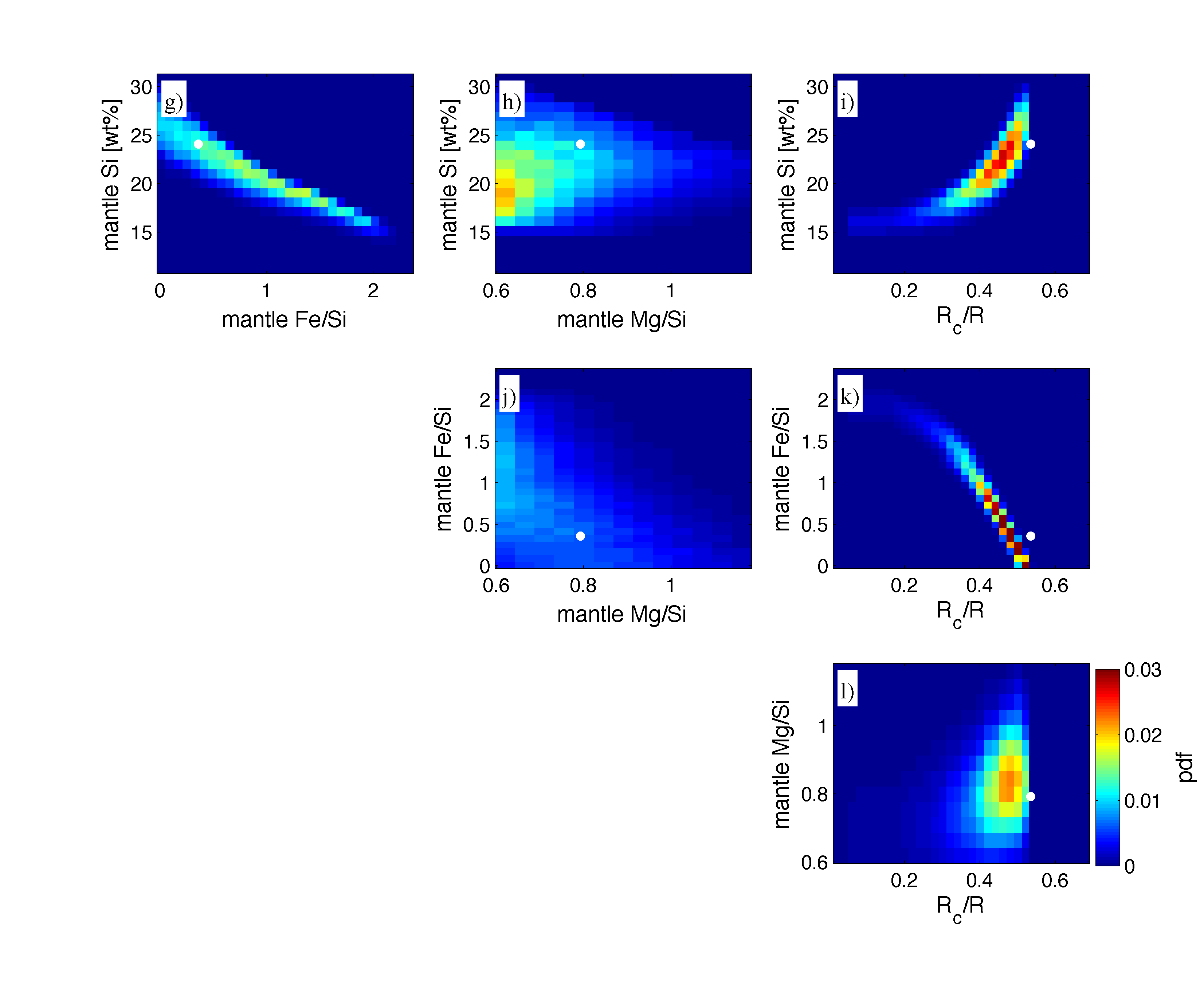}
 \caption{Sampled 2D marginal posterior distributions for the Earth (case A: plots a--f and case B: plots g--l) showing
correlation between parameters $\si$, $\fesima$, $\mgsima$, and core radius $R_c$ (corresponding to Figure \ref{figure_earth}). Independent estimates (Table \ref{table_true}) are marked in white. \label{corr_earth}}
\end{figure*}

Figure \ref{corr_earth} shows the correlation between individual model parameters for case A and B. In contrast to case A, strong correlations in case B are evident between model parameters: $\si$, $\fesima$, and $R_{\rm c}$. This is due to the constraint provided by $\fesi$ that couples core and mantle, i.e., a larger core must be compensated by a lower $\fesima$ or a larger $\si$. No significant correlation of $\mgsima$ to the other model parameters is observed, because data are not able to constrain $\mgsima$ well.

\begin{figure*}[ht]
\center
 \includegraphics[width = 1\textwidth, trim = 5cm 0cm 13cm 0cm, clip]{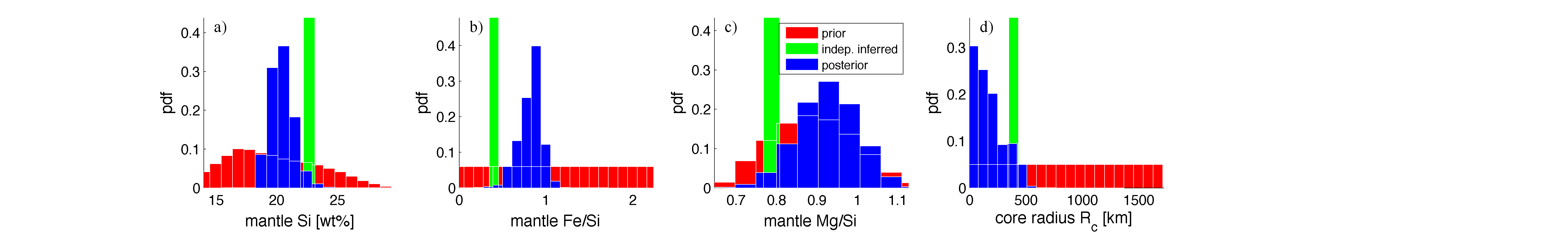}
 \caption{Sampled 1D marginal posterior distributions for the Moon, case B: prior (red), posterior (blue) and independent estimates (green) of model parameters: (a) mantle Si-content $\si$, (b) mantle $\fesima$, (c) mantle $\mgsima$, and (d) core radius $R_c$. Independent estimates are listed in table \ref{table_true}. \label{figure_moon}}
\end{figure*}

\paragraph{Moon} 
Results for the Moon are shown in Figure \ref{figure_moon}.  As for Earth, we see that $\si$, $\fesima$, and $R_{\rm c}$ are well-constrained, i.e. prior and posterior pdf significantly differ, except for  $\mgsima$. Predicted core size matches the independent estimate (Table \ref{table_true}) reasonably well with $0 \le R_{\rm c} \le 380$ km (95\% credible interval), which incidentally also assumes a pure iron core. The mantle composition appears not to be fully retrieved here. This is because the Moon has the lowest $\fesi$ among the terrestrial planets, including chondrites and satellites of the outer solar system \citep{kuskov}. Therefore, the assumption that $\fesi = \fesistar$ leads us to overestimate $\fesima$ and underestimate $\si$. The independent estimate of mantle compositions nonetheless lies within the range of inferred posterior models.

\begin{figure*}[ht]
\center
 \includegraphics[width = 1\textwidth, trim = 5cm 0cm 13cm 0cm, clip]{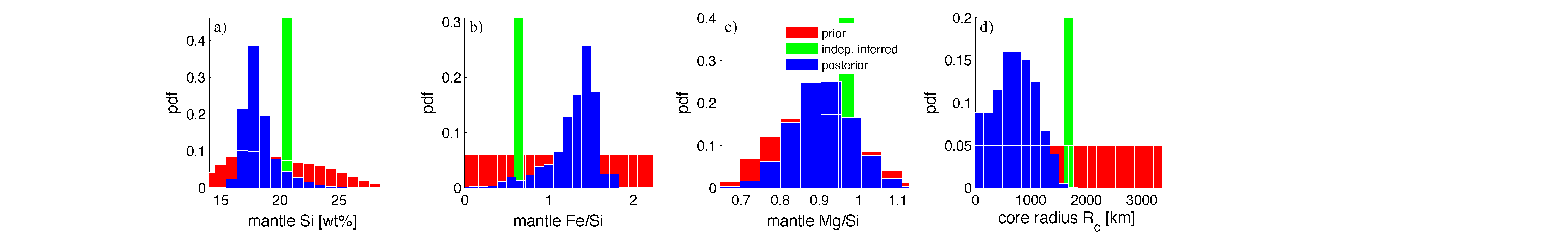}
 \caption{Sampled 1D marginal posterior distributions for Mars, case B: prior (red), posterior (blue) and independent estimates (green) of model parameters: (a) mantle Si-content $\si$, (b) mantle $\fesima$, (c) mantle $\mgsima$, and (d) core radius $R_c$. Independent estimates are listed in table \ref{table_true}. \label{figure_mars}}
\end{figure*}

\paragraph{Mars}
Results for Mars are shown in Figure \ref{figure_mars}. Apart from $\mgsima$, mantle composition ($\si$ and $\fesima$) and core radius ($R_{\rm c}$) appear to be well-constrained. Although independent geophysical estimates of \citet{khan08} lie within the posterior distribution, there is discrepancy that relates to the assumption of sulfur in the core model of \citet{khan08}. As a consequence, we underestimate core size to $0 \le R_{\rm c} \le 1330$ km (95\% credible interval), which leads to an overestimate of $\fesima$ and an underestimate of $\si$. Comparing our results to models  with pure iron cores \citep{rivoldini,khan08}, reduces discrepancy.

\begin{figure*}[ht]
\center
 \includegraphics[width = 1\textwidth, trim = 5cm 0cm 12.7cm 0cm, clip]{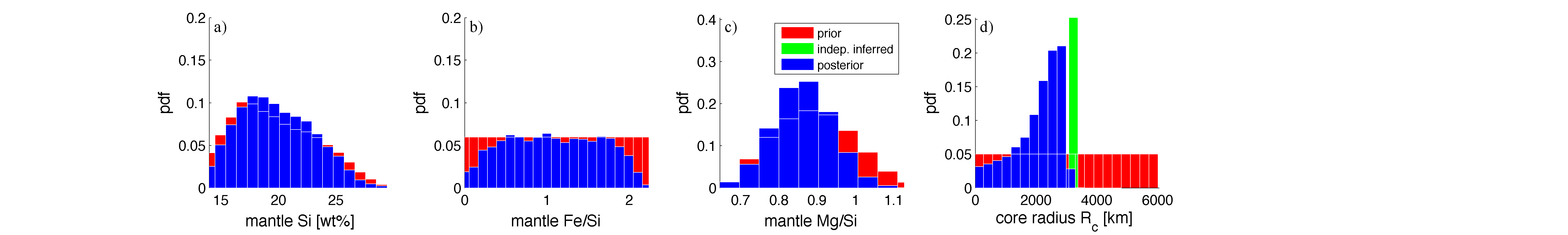}
 \caption{Sampled 1D marginal posterior distributions for Venus, case B: prior (red), posterior (blue) and independent estimates (green) of model parameters: (a) mantle Si-content $\si$, (b) mantle $\fesima$, (c) mantle $\mgsima$, and (d) core radius $R_c$. Independent estimates are listed in table \ref{table_true}. \label{figure_venus}}
\end{figure*}

\paragraph{Venus}
Detailed independent analyses of the mantle composition of Venus are not yet available. \citet[][]{aitta}, e.g. impose  an  Earth-like mantle composition and find that the composition of Venus' core is likely enriched in lighter elements above that of Earth's core to fit mass and radius. Consequently, as shown in Figure \ref{figure_venus}, we underestimate core radius to $0 \le R_{\rm c} \le 2810$ km (95\% credible interval) compared to the independent estimates of \citet{aitta}. Generally, $R_{\rm c}$ is well constrained, while the mantle composition ($\si$, $\fesima$ and $\mgsima$) is only moderately-to-weakly constrained.

\paragraph{Mercury}

\begin{figure*}[ht]
\center
 \includegraphics[width = 1\textwidth, trim = 5cm 0cm 13cm 0cm, clip]{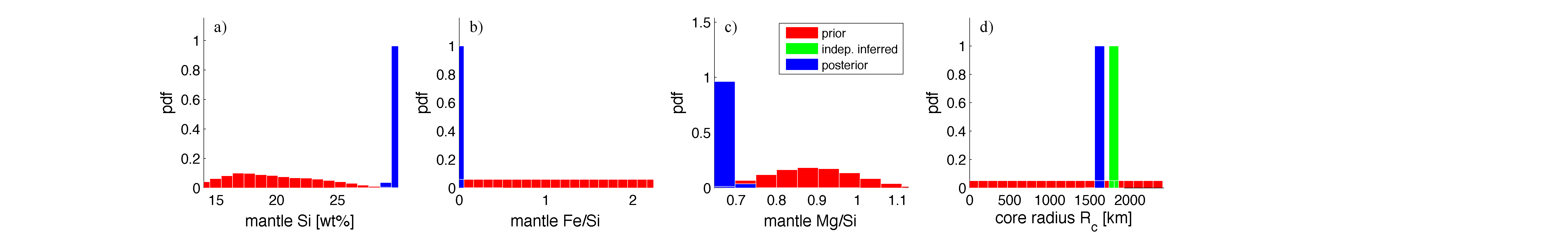}
 \caption{Sampled 1D marginal posterior distributions for Mercury, case B: prior (red), posterior (blue) and independent estimates (green) of model parameters: (a) mantle Si-content $\si$, (b) mantle $\fesima$, (c) mantle $\mgsima$, and (d) core radius $R_c$. No interior model fits the data, i.e., likelihoods are nearly zero. Independent estimates are listed in table \ref{table_true}. \label{figure_mercury}}
\end{figure*}

Mercury has a high bulk density and is thought to have experienced a different accretion scenario or undergone a different post-formation process in comparison to the other terrestrial planets \citep{cameron, strom}. An early giant impact, is likely to have stripped off a large part of its silicate crust and mantle \citep{Benz}, leaving behind a planet with a large iron core. Hence, the constraint $\fesi = \fesistar$ is not applicable to bulk Mercury as a consequence of which it is not possible to match data. 
We nonetheless show the posterior model parameter pdfs in Figure \ref{figure_mercury}, but emphasize that these models have near-zero likelihoods. To improve fit to $M$ a larger iron core is required. This, however, would imply a higher mantle 
Si-content ($\si$) in order to simultaneously match the $\fesi$ constraint, but would move $\si$ outside the prior range.
\\

In summary, the addition of $\fesi$ as constraint (case B) is clearly the key parameter that allows us to: (1) constrain mantle composition, (2) obtain an improved estimate of core size, and (3) significantly reduce model variability. This ultimately arises because the compositional constraint ($\fesi$) results in a strong correlation of the model parameters $\si$, $\fesima$ and $R_{\rm c}$ in the inversion as clearly demonstrated in Figure \ref{corr_earth}. 
The data are not able to constrain $\mgsima$, since $\mgsima$ appears not to be significantly correlated to the other model parameters.
We find that independent estimates of $\si$, $\fesima$ and $R_{\rm c}$ are relatively well predicted by our method for Earth, Moon, Mars, and Venus. This is not the case for Mercury, where no model is found that fits data given our model assumptions.

\section{Results}
\label{Results}
 
\subsection{Application to synthetic exoplanets} 
\label{synthexo}
In this section, we apply our method to a set of synthetic planets where mass and radius uncertainties are considered to be artificially small in order to demonstrate its performance. We consider a range of synthetic exoplanets between $0.1 M_E<M<10 M_E$ and $0.5 R_E<R<1.7 R_E$ for both case A (inversion of mass and radius) and B (inversion of mass, radius, and bulk abundance constraints). Prior model ranges are listed in Table \ref{table_range}.
 We employ the same standard deviations used in the case of the terrestrial solar system planets, i.e., $\sigma_{M} = 0.01 M$, $\sigma_{R} = 0$, and $\fesi = \fesistar = 1.69 \pm 0.18$ \citep{lodders03}, as well as fixing the minor components (Na$_2$O, CaO, and Al$_2$O$_3$) to chondritic values \citep{lodders03}.


\begin{figure*}[ht]
\center
 \includegraphics[width = 1\textwidth, trim = 1cm 0cm 1cm 0cm, clip]{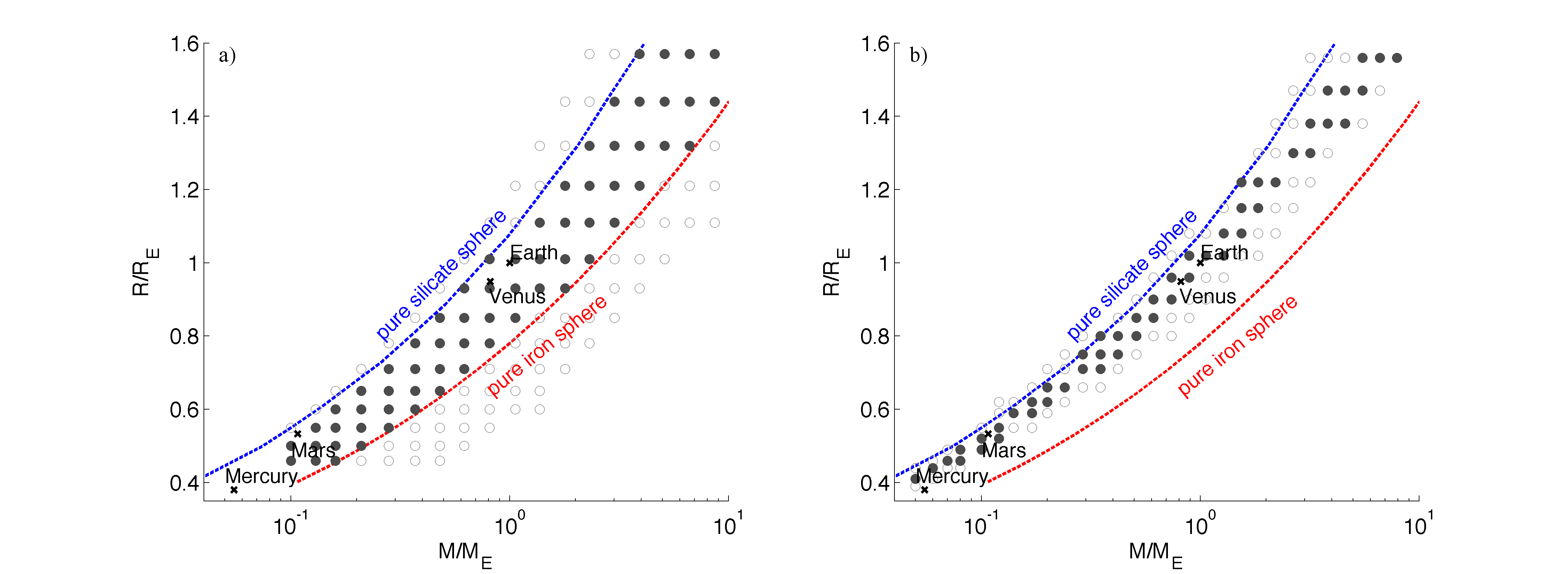}
 \caption{Range of synthetic rocky planets that fit data (black dots) of (a) $M$ and $R$ (case A), and (b) $M$, $R$, and stellar abundance constraints (case B). Mass-radius curves for pure iron (red) and pure silicate spheres (blue, zero amount of FeO) were calculated independently. In case A, all synthetic planets that fit data (black dots) fall within the limits of a pure silicate (blue line) and pure iron sphere (red line). In case B, this region is reduced to a smaller area indicating that chemical bulk ratio ($\fesi$) introduces strong additional constraints on the internal structure. Synthetic planets for which no internal structure could be derived within model assumptions (pure iron core surrounded by a rocky mantle) are indicated by gray open circles. Note that for every circle in the plots an inverse problem has been solved. See main text for further details. \label{figure_scatter}}
\end{figure*}

Figure \ref{figure_scatter} shows the range of synthetic rocky planets that fit data for case A (left panels) and B (right panels). Note that for every synthetic planet we solved an inverse problem, i.e., for every point in the $M$-$R$-plane in Figure \ref{figure_scatter}, a full inverse problem is solved yielding separate pdfs for all parameters. 
When inverting only $M$ and $R$ (case A), all planets are seen to fall within the extreme limits of a pure iron and a pure silicate sphere (Figure \ref{figure_scatter}) as expected. 
 By adding the compositional constraint ($\fesi$) inferred from observations of the host star (here the Sun), the extreme case of a pure iron sphere, for example, is clearly unable to satisfy this constraint. As a consequence, for case B synthetic planets  plot within a smaller range than for case A. Mercury, for example, lies outside the permissible region for reasons discussed above.

With our assumption that planets are composed of a pure iron core surrounded by a rocky mantle of arbitrary size, it is not surprising that internal structures could only be derived for planets with structures lying between a pure iron and a pure rocky sphere. By adding a compositional constraint ($\fesi$) the range of possible internal structures is sharply reduced, leading to exclusion of Mercury-type planets.


\begin{figure*}[ht]
\center
 \includegraphics[width = 1\textwidth, trim = 1cm 0cm 6cm 0cm, clip]{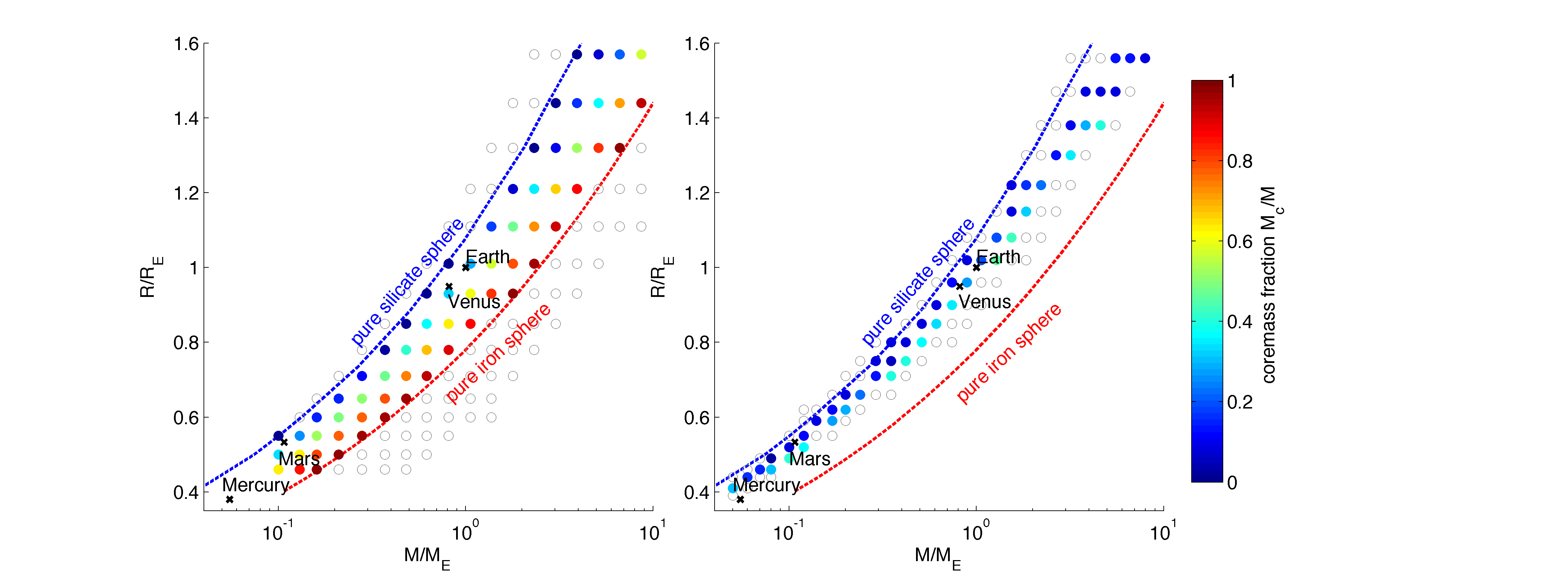}
 \caption{Mean posterior core mass fraction $M_c/M$ plotted for the range of synthetic rocky planets that fit (a) $M$ and $R$ (case A), and (b) $M$, $R$, and stellar abundance constraints (case B). (See also Figure \ref{figure_scatter}.) \label{coremass}}

\end{figure*}

Between the limits of a pure iron and pure silicate sphere, bulk densities decrease towards the silicate-sphere limit. Thus, we also expect  core size to decrease towards the silicate-sphere limit. To verify that this trend is indeed obtained here, we  
plot inverted core mass fraction $M_{\rm c}/M$  in Figure \ref{coremass} for all considered synthetic exoplanet cases shown in Figure \ref{figure_scatter}.  In the limit of a pure iron (silicate) sphere $M_{\rm c}/M \rightarrow$ 1 (0). 
For case B, models with $M_{\rm c}/M \ge 0.6$, i.e., large iron cores, are evidently excluded as it becomes difficult to increase mantle Si-content in order to satisfy the compositional constraint ($\fesi$).

 In order to quantify the information content of data, i.e. the degree to which data constrain internal structure, we compute the Shannon entropy measure ($\shann$) \citep{tarantola}, which allows comparison of prior and posterior pdfs. If $\shann \rightarrow 1$, data are only weakly constraining models, whereas if $\shann \rightarrow 0$, data pose strong constraints on the models (see Appendix \ref{shannon} for details).
Figure \ref{figure_shannon} shows $\shann$ for all inverted model parameters. 
For case A terrestrial-type planets, we find that core radius $R_{\rm c}$ is generally well-constrained, whereas mantle composition is weakly constrained. For $\si$, $\fesima$, and $\mgsima$ the relative Shannon entropy measures are larger than 0.8 throughout the tested range of synthetic planets. There are few planets that have bulk densities that are very close to the pure silicate sphere for which $\fesima$ appears to be better constrained. Generally for case A, the inversion scheme is not sensitive to variations in mantle  composition but mostly sensitive to variations in core radius $R_{\rm c}$. Stated differently, any variations in mantle composition can be compensated by tiny variations in core radius, while variations in core radius must be compensated by large changes in mantle composition in order that data ($M$ and $R$) be matched. 

From Figure \ref{figure_shannon} we observe the following trend for case A: $R_{\rm c}$ appears to be better constrained for  smaller and denser planets.
The denser a rocky planet is, the more it is dominated by its core and the less mantle material is available whose compositional changes could compensate for variations in core radius. Similarly, the smaller a planet is, the less absolute mantle mass is available to compensate for a higher variability in core radius.
Therefore, the precision with which core radius can be determined rests ultimately on the available mantle mass. 

For case B, the addition of $\fesi$ significantly reduces model variability, i.e. correlation between model parameters is increased. $R_{\rm c}$, $\si$ and $\fesima$ are better constrained compared with case A, while $\mgsima$ is only weakly constrained as for case A. Note however, that the prior bounds $\mgsima$ are significantly narrower for case B than for case A, because of the stellar abundance constraint.
The relative Shannon entropy measures for core radius $R_{\rm c}$ in case B are similar but slightly smaller than for case A. 

As discussed for the terrestrial solar system planets, only the core size can be constrained in the case of exoplanets for which only mass and radius are known (case A).  In case B, the compositional constraint ($\fesi$) introduces a strong correlation between the model parameters $\si$, $\fesima$ and $R_{\rm c}$.  This enables us to constrain mantle composition.

\begin{figure*}[ht]
\center
 \includegraphics[width = 1\textwidth, trim = 2cm 5cm 1cm 2cm, clip]{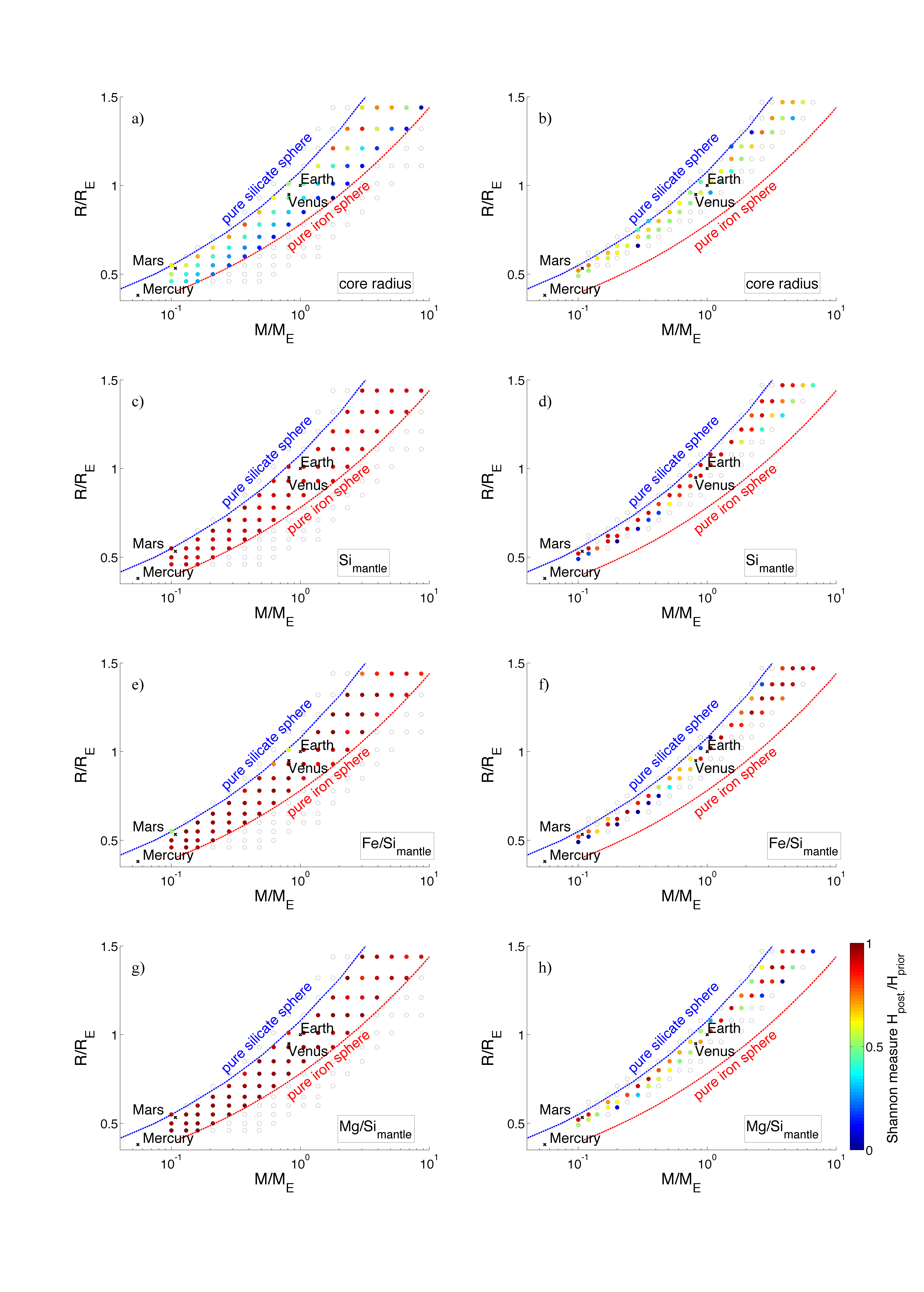}
 \caption{Relative Shannon entropy measure $\shann$ of model parameters plotted for the range of synthetic rocky planets that fit a given $M$ and $R$ (case A, left panels), and $M$, $R$, and $\fesi$ (case B, right panels). Model parameters are well constrained for $\shann \rightarrow 0$ (blue colors) and not constrained for $\shann \rightarrow 1$ (red colors). For case B, variations of $\shann$ along the grid appear not to be smooth which is due to the poor grid resolution. See also Figure \ref{figure_scatter}. \label{figure_shannon}}
\end{figure*}

\subsection{Application to confirmed exoplanets}
\label{confirmedexo}
\begin{figure*}[ht]
\center
 \includegraphics[width = 1\textwidth, trim = 2cm 0cm 2cm 0cm, clip]{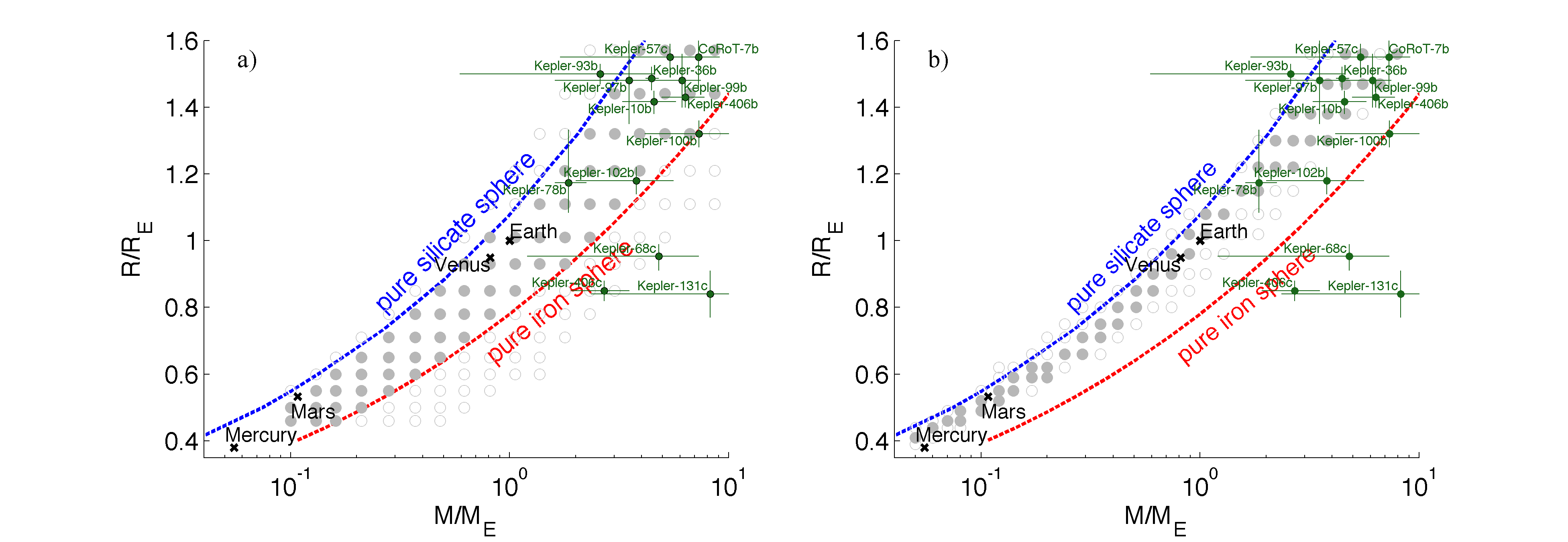}
 \caption{Confirmed exoplanets plotted against Figure \ref{figure_scatter}: (a) Case A and (b) case B. \label{figure_known}} 
\end{figure*}

In this section we apply our methodology to actual exoplanet measurements in order to derive models of the internal structure. At this point in the development of our approach we are still limited by the assumptions made for the overall structure (iron core and rocky mantle with no ice and no atmosphere) and hence this application serves more as an illustration of the overall capabilities of our method. 
In Figure \ref{figure_known} we plot all confirmed exoplanets in the mass-radius range of interest on top of our synthetic exoplanet calculations (Figure \ref{figure_scatter}). Masses, radii, and uncertainty ranges of the observed exoplanets are listed in Table \ref{table_exo}. Kepler-68c, Kepler-131c and Kepler-406c have estimated mean densities above that of pure iron and clearly present special exotic cases. Kepler-36b has been chosen to demonstrate the application of our method to actual observations because of small uncertainties on mass and radius. To our knowledge, no stellar elemental abundances of $\fesistar$ and $\mgsistar$ of its host star other than for the Sun are available. Hence, we test case B assuming solar elemental abundances. Since the Kepler mission has been targeting Sun-like stars \citep{borucki}, this is a permissible assumption. 

\begin{figure*}[ht]
\center
 \includegraphics[width = 1\textwidth, trim = 5cm 0cm 13cm 0cm, clip]{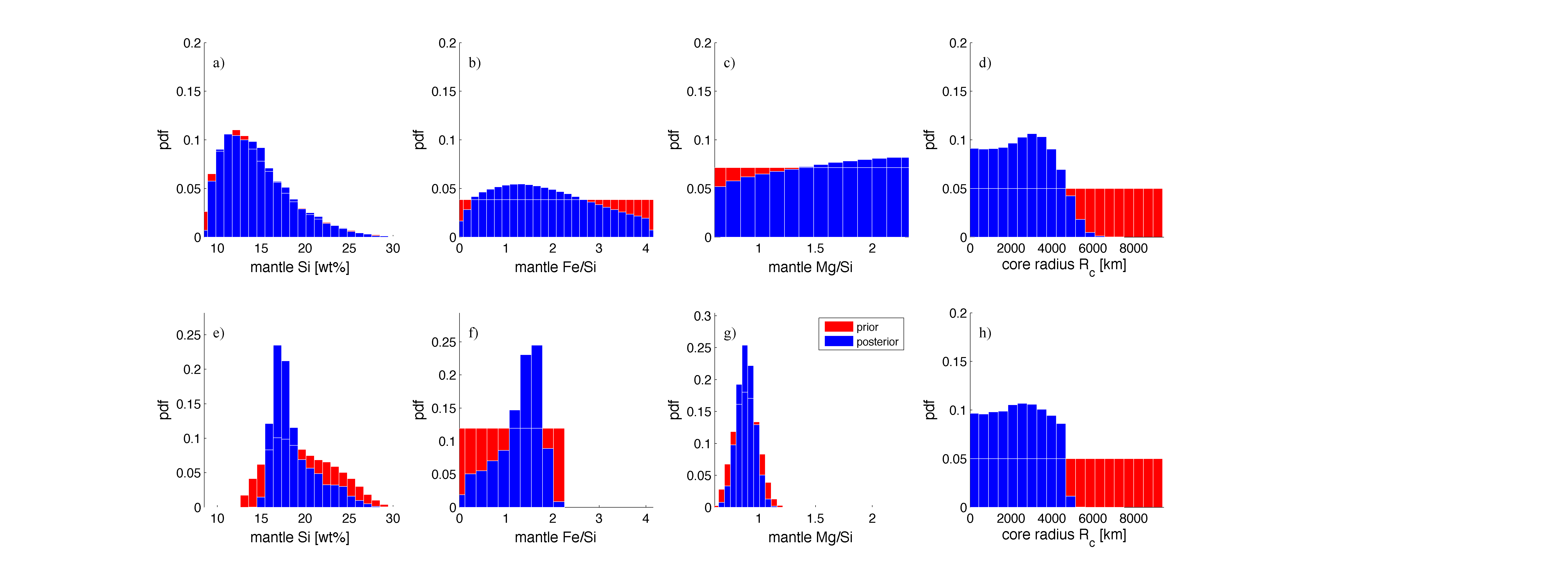}
 \caption{Sampled 1D marginal posterior distributions for Kepler-36b, case A (a-d) and B (e-h): prior (red) and posterior (blue) of model parameters: (a,e) mantle Si-content $\si$, (b,f) mantle $\fesima$, (c,g) mantle $\mgsima$, and (d,h) core radius $R_c$. Data uncertainties are $\sigma_{\fesi} = 10\%$, $\sigma_{R} = 2\%$, and $\sigma_{M} = 7\%$. \label{K36}}
\end{figure*}

Figure \ref{K36} shows the results for Kepler-36b. 
For case A, we see that the mantle ratios $\fesima$ and $\mgsima$ are not constrained at all, while $\si$ and $R_{\rm c}$ are moderately and well constrained, respectively. By adding $\fesi$ to the data (case B), $\si$ and $\fesima$ are significantly better constrained. Thus, incorporating observations of stellar elemental abundances has a big impact on our ability to constrain model parameters. 
The posterior range of $R_{\rm c}$ is very large in both cases ranging from zero to about half the planet radius. 
We argue that mass and radius uncertainties of 7 \% and 2 \%, respectively, are the reason for such a large spread. 

The example of Kepler-36b illustrates the necessity of having well-measured masses and radii in order to derive constraints on the interior structure. Furthermore, it exemplifies the large impact of elemental abundance ratios in order to constrain planetary interior structure. Since it will be impossible to directly measure these, spectroscopically-determined values measured from the photosphere of the host star should serve as proxies.

\begin{center}
\begin{table*}[ht]
\caption{Masses and radii of observed exoplanets \label{table_exo}}
\begin{center}
\begin{tabular}{lcccc}
\hline\noalign{\smallskip}
 planet & radius $R/R_E$ & mass $M/M_E$ & reference & mean bulk density $\rm g/cm^3$ \\
\noalign{\smallskip}
\hline\noalign{\smallskip}
CoRoT-7b & 1.55 $\pm 0.10$ 	 & 7.31 $\pm 1.21$ & \cite{moutou}			& 10.82 \\
Kepler-10b &1.416 $\pm 0.033$ 	 & $4.56 ^{+1.17}_{-1.29} $ & \cite{batalha}   &8.89 \\
Kepler-36b & 1.486 $\pm 0.035 $ & 	$ 4.45 ^{+0.33}_{ -0.27}$& \cite{carter}	&7.48 \\
Kepler-57c &1.55 $\pm 0.04 $	 & $ 5.4 \pm 3.7		$& \cite{steffen}	&8.00 \\
Kepler-68c & $0.953^{+0.037}_{-0.042}$ & $4.8 ^{+2.5 }_{-3.6}	$& \cite{gilliland}	&30.58 \\
Kepler-78b & $1.173^{+0.159}_{ -0.089} $ & $1.86 ^{+0.38 }_{-0.25}$& \cite{pepe}     &6.35 \\
Kepler-93b & 1.50 $\pm 0.03 $  & $2.59 \pm 2.00	$		& \cite{weiss11}	&4.23 \\
Kepler-97b & 1.48 $\pm 0.13 $  & $3.51 \pm 1.90$		& \cite{weiss11}		&5.97 \\
Kepler-99b & 1.48 $\pm 0.08 $  & $6.15 \pm 1.30  $	& \cite{weiss11}			&10.46 \\
Kepler-100b & 1.32 $\pm 0.04 $ & $7.34 \pm 3.20 $		& \cite{weiss11}		&17.60 \\
Kepler-102b & 1.18 $\pm 0.04 $ & $3.8 \pm 1.8	  $	& \cite{weiss11}			&12.75 \\
Kepler-131c & 0.84 $\pm 0.07$ & $8.25 \pm 5.90	$	& \cite{weiss11}		& 76.74\\
Kepler-406b & 1.43 $\pm 0.03 $ & $ 6.35 \pm 1.40 $ & \cite{weiss11}			&11.97 \\
Kepler-406c & 0.85 $\pm 0.03$ & $2.71 \pm 0.80 $ & \cite{weiss11}			& 24.33 \\
\hline
\end{tabular}
\end{center}
\end{table*}
\end{center}

\subsubsection{Information content of the data}

The limiting factor on our ability to constrain interior structure is measurement precision. In this section, we study in more detail the link between measurement precision and resulting constraints on internal structure.

\begin{figure*}[ht]
\center
 \includegraphics[width = 1\textwidth, trim = 2cm 0cm 2cm 0cm, clip]{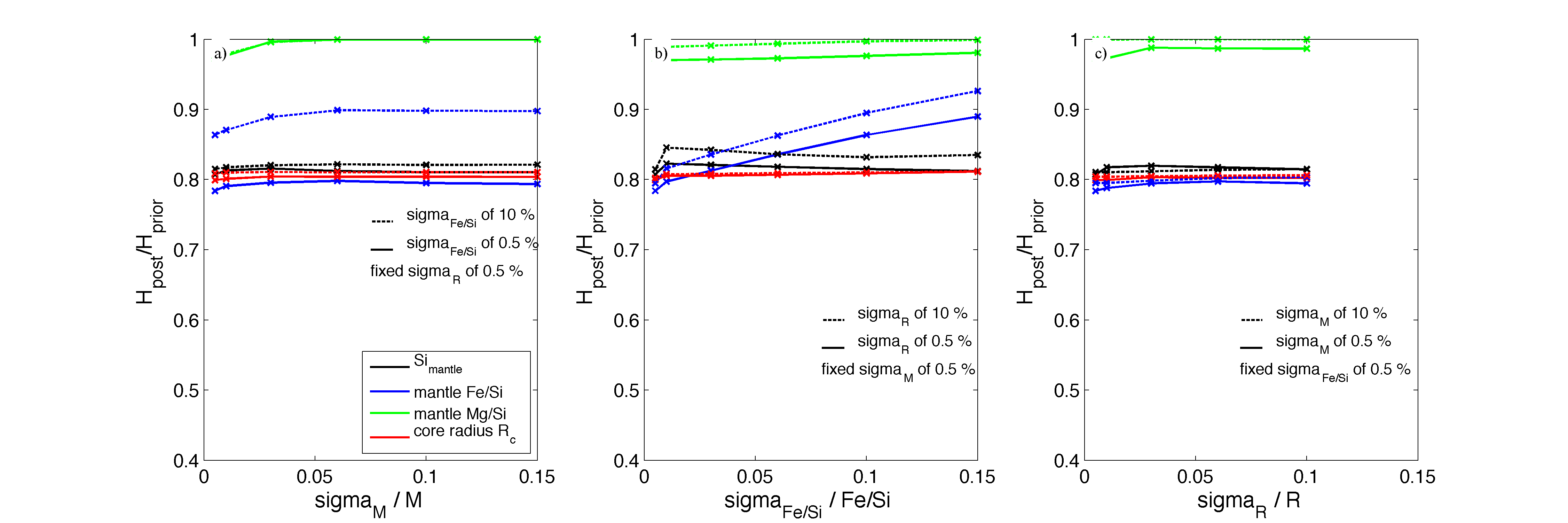}
 \caption{Summary of information content for Kepler 36-b ($M = 4.45 M_E$, $R=1.486 R_E$): Relative Shannon measure as a function of (a) $\sigma_{M}$ while fixing $\sigma_{R}$ at 0.5\% (b) $\sigma_{\fesi}$ while fixing $\sigma_{M}$ at 0.5\% (c) $\sigma_{R}$ while fixing $\sigma_{\fesi}$ at 0.5\%. \label{uncertK}} 
\end{figure*}

\begin{figure*}[ht]
\center
 \includegraphics[width = 1\textwidth, trim = 2cm 0cm 2cm 0cm, clip]{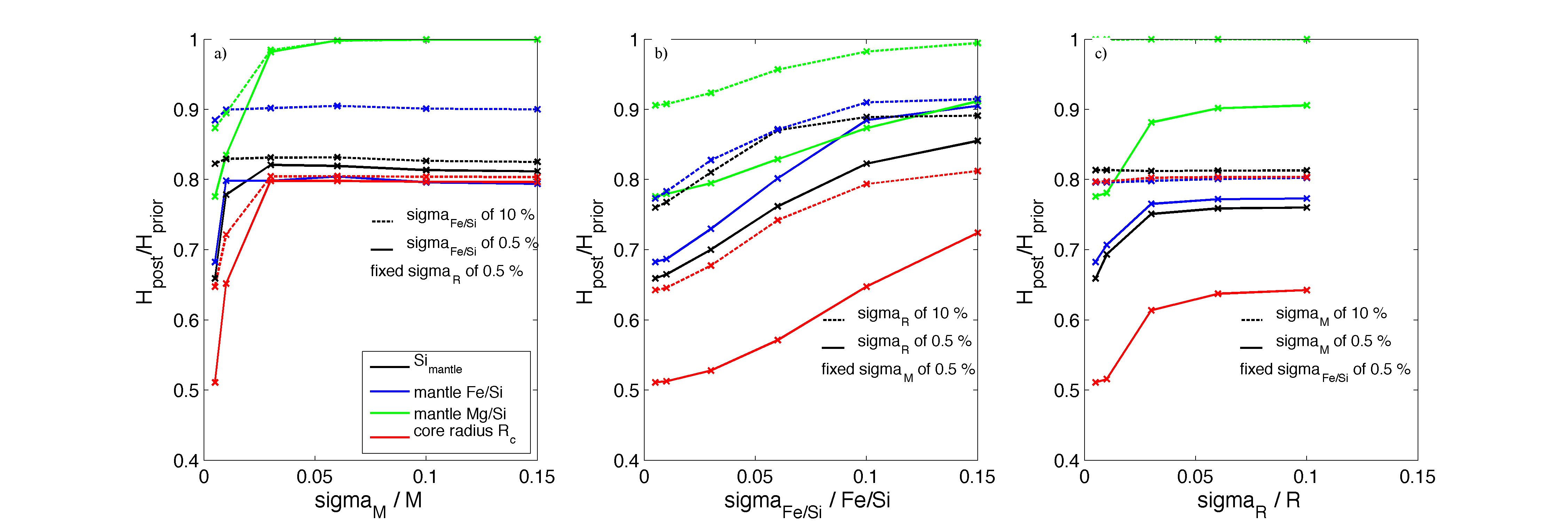}
 \caption{Summary of information content for an Earth-sized body ($M = 1 M_E$, $R=1 R_E$): Relative Shannon measure as a function of (a) $\sigma_{M}$ while fixing $\sigma_{R}$ at 0.5\% (b) $\sigma_{\fesi}$ while fixing $\sigma_{M}$ at 0.5\% (c) $\sigma_{R}$ while fixing $\sigma_{\fesi}$ at 0.5\%. \label{uncertE}} 
\end{figure*}

For this, we compute the relative Shannon entropy measure $\shann$ for each model parameter for a range of measurement uncertainty. In these calculations, inversions were performed using a grid-search method.
The results are shown in Figure \ref{uncertK} for the example of Kepler-36b and in Figure \ref{uncertE} for an Earth-like planet. The results can be summarized as follows:
\begin{itemize}
\item Overall, a decrease in data uncertainty leads to a smaller $\shann$ and thus better constrained parameters. However, for large uncertainties $\shann$ is nearly constant and relatively large.

\item Uncertainties on $M$, $R$, and $\fesi$ appear equally important.

\item The inherent degeneracy of the problem is limiting our ability to constrain interior structure even in the case of very small data uncertainties. For the cases shown, $\shann$ does not fall below 0.5.

\item Our ability to constrain interior structure is significantly different for larger and denser exoplanets such as Kepler-36b compared to Earth-like bodies. As shown for Kepler-36b, $\shann$ decreases only at extremely small data uncertainties whereas for an Earth-like body,  $\shann$ decreases significantly earlier. Given their large difference in mass, this indicates that measurements with a given precision are able to better constrain the internal structure of small-mass planets than large-mass planets. Large planets have more absolute mantle mass available whose compositional changes are able to compensate for a higher variability in core radius, ultimately  reducing our ability to constrain interior models.

\item Over the entire tested range of uncertainties on $\fesi$ (0-15\%), significant improvements in our ability to constrain interior structure can be observed, whereas for mass and radius this is only evident in the case of small uncertainty ranges ($ < 5\%$). 
\end{itemize}

\section{Discussion}
\label{Discussion}

The results presented here have to be interpreted in the light of the assumptions made. Firstly, we consider terrestrial-like planets that consist only of silicate mantles and pure iron cores. \citet{sotin07} showed that rocky planets can be described with great accuracy by only using the four elements Si, Mg, Fe, O. Secondly, we assume a pure iron composition for the core. Volatiles (H, He, H$_2$O, etc.) are currently neglected. 
Compared to Earth, where  the core include lighter elements (e.g., S, Si, C, O), our method therefore systematically overestimates core density and underestimates core radius. This introduces biases in the estimates of mantle composition due to the strong correlations between mantle composition and core radius. In the future, lighter elements (e.g., Si) will be included in the core. Thirdly, our approach assumes subsolidus conditions and perfectly known Equation-of-State parameters for all considered compositions. For exoplanets that are larger than Earth, extrapolations of parameters to pressures and temperatures beyond those measured in the laboratory are required. This introduces uncertainty in the EoS modeling that we will address in the future.

While the mean density of a terrestrial-like planets can be derived from measurements of mass and radius, its internal structure is only relatively poorly constrained by these two quantities as also shown by \citet[e.g.,][]{rogers2010}. We find that adding $\fesi$ and $\mgsi$ as compositional constraints significantly improves this situation. 
These additional constraints significantly  narrow the prior range of mantle compositions \citep[for details on use of different priors see e.g.,][]{rogers2010}.
   In particular,  $\fesi$ introduces a strong correlation between model parameters $\si$, $\fesima$, and $R_{\rm c}$, which significantly reduces the range of possible models. Hence, $\fesi$ plays a key role in constraining mantle composition and internal structure.

Our ability to constrain interior structure is directly limited by data uncertainties. We have demonstrated that uncertainties on mass, radius, and elemental abundances appear to be equally important.
 While it is worth striving to increase the measurement precision on all input data, the knowledge of $\fesistar$ and $\mgsistar$ is particularly important as it represents a relatively powerful means of reducing the degeneracy of possible interior models. 
The level of inherent degeneracy in model solutions actually depends on the specific values of $M$ and $R$ (as discussed for Figure \ref{figure_shannon}) \citep{rogers2010}: for given measurement accuracy of mass and radius, interior structure is better constrained for an Earth-sized planet than for a larger body (e.g., Kepler-36b).

In an earlier study, \citet{rogers2010} employed Bayesian analysis to study exoplanet GJ 581d in an attempt to constrain core mass and silicate mantle mass fractions of MgSiO$_{3}$ and FeSiO$_{3}$.
Based on mass and radius alone, \citet{rogers2010} also concluded that an exact interior solution cannot be inferred since the problem is highly underdetermined.
However, \citet{rogers2010} argue against the use of bulk abundance constraints inferred from $\fesistar$ and $\mgsistar$ in order to remain as independent as possible of planet formation models. 
It should be noted that if \citet{rogers2010} were to use bulk abundance constraints core size and mantle composition would be fixed since they use a parameterized phase-diagram approach and restrict the mantle mineralogy to Mg$_{1-\chi}$Fe$_{\chi}$SiO$_{3}$. Consequently, the problem would no longer be degenerate. Clearly, our approach is more general because it allows for arbitrary elemental abundance ratios while using stellar abundance constraints as a means of further reducing model variability.

With regard to planet formation, these studies suggest that some interior compositions are more likely than others. 
Recent planet formation studies accounting for equilibrium condensation of material \citep[e.g.,][]{thiabaud} have shown that most planets have a bulk refractory  (Fe, Si, and Mg) composition that appears to be indistinguishable from that of the host star because species of Fe, Si and Mg condense at similar temperatures ($\sim$ 1000 K). Hence, most planets will have similar refractory elemental ratios as those in the original disc which is assumed to be well-represented by stellar ratios.
Also, O, Fe, Si, and Mg are predicted to be among the most frequent
solid species in circumstellar disks \citep[e.g.,][]{elser, gilli} as a consequence of which models with large iron cores
and small silicate mantles are generally unlikely to form.
 Based on these observations \citet{valencia07a} also proposed the general use of a minimum bulk value for Fe/Si. Here, we have followed \citet{grasset09}, who considered bulk Mg/Si and Fe/Si to be dictated by their abundance in the host star's photosphere and adopt these as proxies for planet bulk composition. 

  For the terrestrial solar system planets, we showed that using solar photospheric abundances ($\fesistar$ and $\mgsistar$) yields tighter constraints on the internal structure, except for Mercury. Mercury's interior is unique among the terrestrial planets and is usually explained by the removal of a large fraction of its silicate mantle by either a giant impact or by evaporation due to the proximity of the planet to the sun \citep{Benz}. Consequently, our scheme is not able to find interior models that fit data for Mercury. In spite of this, we argue for the use of stellar abundance constraints, because of the ability of our scheme to predict for which planets alternative accretion scenarios or post-formation processes must be envisaged. We also note that meteorites, which are assumed to be the building blocks of the terrestrial planets, have abundances of refractory elements within 10\% of solar values \citep{lodders03}.

\section{Conclusions and Outlook}
\label{Conclusions}

We have presented a Bayesian inversion method in order to constrain the interior structure of rocky exoplanets from measurements of mass and radius. Since stellar refractory elemental abundances are likely to be generally a good proxy for planetary bulk compositions, they represent powerful additional constraints allowing us to significantly improve determination of internal structure.
Our proposed scheme is more general than previous works on mass-radius relationship of rocky exoplanets, in that we obtain confidence regions of interior structures of general mantel composition and core structure. Furthermore, we investigated how interior structure models depend on various parameters in the form of prior information, data, and data uncertainties.

We have applied the method to the terrestrial planets of our solar system and compared the results to independent estimates and, except for Mercury, find generally good agreement.
Following this, we have applied our method to synthetic exoplanets in a large mass and radius range as well as to exoplanet Kepler-36b. We have investigated two different cases: case A where we invert mass and radius, and case B where we additionally consider stellar elemental abundances obtained from photospheric observations of the host star. 
 We summarize our findings in the following:

\begin{itemize}

\item Photospheric elemental abundances of the host star ($\fesistar$ and $\mgsistar$) are key parameters in reducing model degeneracy through the introduction of correlations between mantle composition and core size.
\\
\item How well mantle composition and core radius can be generally constrained depends on data and data uncertainties. There is an inherent degeneracy that limits our ability to constrain the interior structure even in the case of small data uncertainties. Independently of these, it seems that model variability depends on mass and density of a planet. Thus, it is our contention that a case-by-case probabilistic inversion that provides model parameter uncertainties is indispensable in order to rigorously characterize interior structure.
\\
\item Measurement uncertainties on mass, radius, and stellar abundance constraints appear to be equally important.
\\
\item For improved characterization of interior structure better estimates of $\fesistar$ and $\mgsistar$ are required. We have shown for Kepler-36b and an Earth-sized planet, that significant improvements on model parameter estimation can be achieved by reducing the uncertainty of $\fesistar$ and $\mgsistar$.

\end{itemize}

Space missions that aim at characterizing exoplanets by means of the transit method seek the most precise measurement of $R$ and hope for follow-up missions to provide precise measurements for $M$. However, since data precision also depends on characteristics of target star and observation time (integration time, number of transits observed), a careful weighing of costs and benefits is crucial for the success of a mission. 
We have demonstrated how the relative Shannon entropy can be used in future missions (e.g., CHEOPS, TESS, PLATO) as a means of optimizing the scientific return.
In particular, we are able to
quantitatively assess the improvement in interior
structure-determination of a specific exoplanet target due to an increase in the measurement
precision of mass and radius.
To illustrate this we consider the case (discussed in Figure \ref{uncertK}), where the predominantly flat curves imply that smaller uncertainties on mass and radius do not provide significantly better insights on the interior structure. For an Earth-sized exoplanet, however, smaller uncertainties on mass and radius can lead to an improved understanding of interior structure (Figure \ref{uncertE}).

An increasing number of Super-Earths are being observed. Super-Earths lie in the intermediate mass-range between terrestrial planets and the gas/ice giants in the solar system with different scenarios for their interiors. Their interior structures (e.g., purely rocky composition or predominantly water/carbon compounds) and formation histories are, however, a matter of debate, as is the question of whether Super-Earths can harbour life \citep[e.g.][]{valencia07c, howe}. 
A potential way of answering these questions is to invert for the physico-chemical structure in order to narrow down the possible types of interior models. 

In future studies, we will extend the dimensionality of our problem to include more elements, e.g., hydrogen and water, so as to model also exoplanets with atmospheres and/or oceans. This will eventually allow us to study how interior types are distributed among stars and what can be learned from such distributions about planet formation.
However, for a given data set, increased model complexity is generally accompanied with a larger degeneracy in model solutions. Additional data would help to restrict model variability. For example, dynamic observations can constrain a planet's tidal dissipation and thereby provide constraints on the distribution of rigidity and density \citep[through the Love number, e.g.,][]{delisle}.

\begin{acknowledgements}

We would like to thank a very enthusiastic referee for comments that led to a much improved manuscript. This work was supported by the Swiss National Foundation under grant 15-144 and the ERC grant 239605. It was in part carried out within the frame of the National Centre for Competence in Research PlanetS. We would like to thank Lauren Cooper for assistance in the early
stages of this project.

\end{acknowledgements}

\appendix

\section{Prior of mantle Si-content}
\label{priorsi}

 The prior bounds on $\si$ depend on the prior bounds of $\fesima$ and $\mgsima$, since the mass fractions of all oxides in the NCFMAS chemical system must sum to one. The mass fraction of Si can be expressed as
$$
\si = \frac{(1-\chi_{\rm Na_2O, CaO, Al_2O_3} )}{(\frac{\mu_{\rm SiO_2}}{\mu_{\rm Si}}+\fesima \frac{\mu_{\rm FeO}}{\mu_{\rm Fe}} + \mgsima \frac{\mu_{\rm MgO}}{\mu_{\rm Mg}} )} 
$$
where $\mu$ are respective molar weights and the mass fraction of minor oxides are combined in $\chi_{\rm Na_2O, CaO, Al_2O_3}$. Transforming from a uniform prior in $\fesima$ and $\mgsima$ to the variable $\si$ we find the prior of $\si$ being
$$
p(\si) \propto  \lvert \frac{\partial \fesima}{\partial \si} \rvert = 
\label{priorsi} \frac{(1-\chi_{\rm Na_2O, CaO, Al_2O_3} )}{\si^2 \frac{\mu_{\rm FeO}}{\mu_{\rm Fe}}}
$$
This implies a non-uniform prior for $\si$. Outside the given interval in Table \ref{table_range} the prior distribution of $\si$ is essentially zero.

\section{Temperature profile}
\label{tempvary}

\begin{figure*}[ht]
\center
 \includegraphics[width = .4\textwidth, trim = 35cm 0cm 5cm 1.1cm, clip]{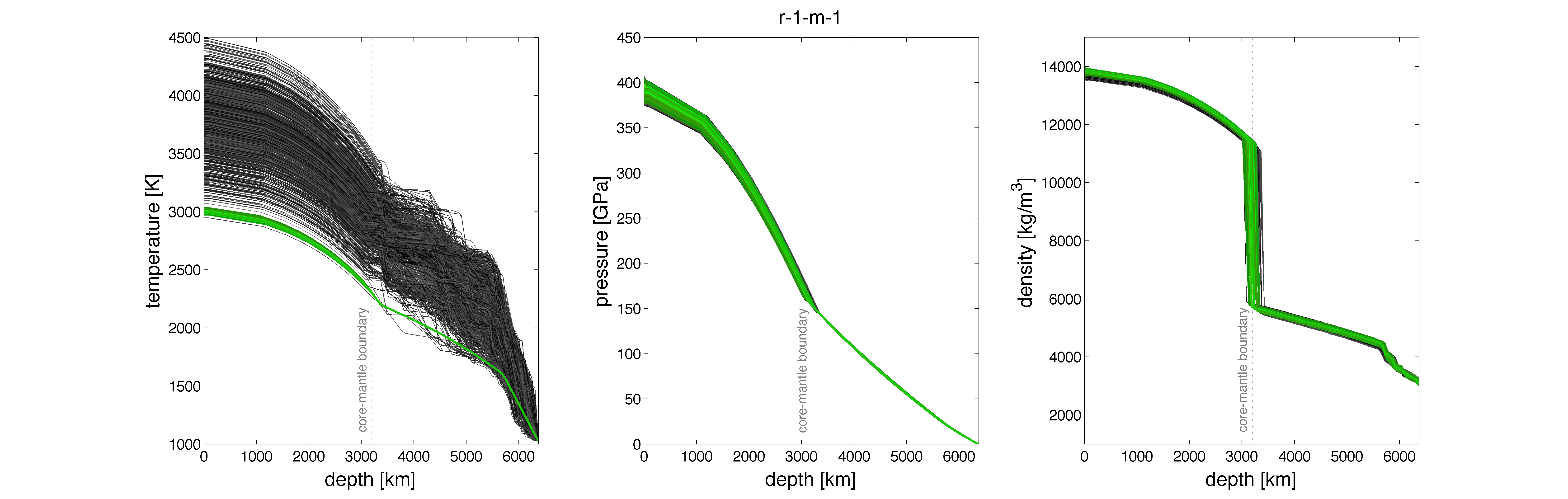}
 \caption{Sampled profiles of density profiles for Earth. A variable mantle temperature (gray profiles) only introduces little additional variation in density compared to profiles with fixed mantle temperature (green colors). \label{figure_temp}}
\end{figure*}

We assume a fixed mantle temperature profile that is based on the Earth model. A variable temperature profile for the mantle only introduces negligible variations in the density profile ($< 1$ \%). This is demonstrated in Figure \ref{figure_temp}. Surface temperature is fixed at 1000 K and although temperature varies between 2100 K to 3500 K at the core mantle boundary, resultant changes in density are relatively small (Figure \ref{figure_temp}).

\section{Silicate model chemical system}
\label{fmssystem}

\begin{figure*}[ht]
\center
 \includegraphics[width = .9\textwidth, trim = 5cm 0cm 13cm 0cm, clip]{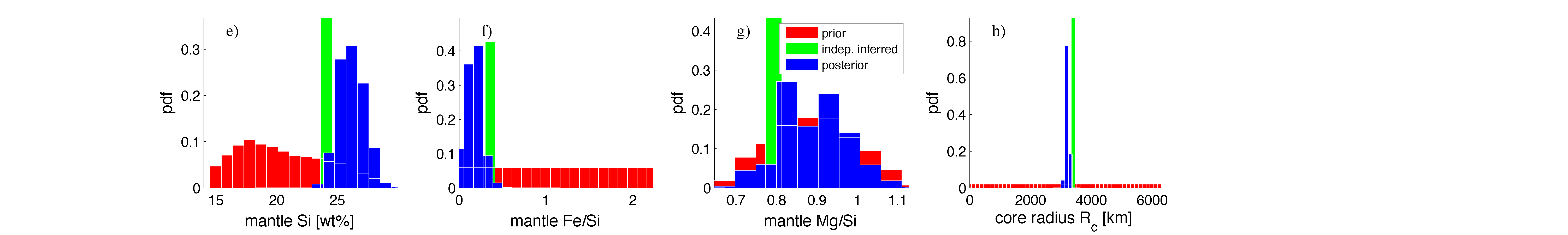}
 \caption{Sampled 1D marginal posterior distributions for Earth using the FMS system: prior (red), posterior (blue) and independent estimates (green) of model parameters: (a) mantle Si-content $\si$, (b) mantle $\fesima$, (c) mantle $\mgsima$, and (d) core radius $R_c$. Independent estimates are listed in table \ref{table_true}. \label{fms_earth}}
\end{figure*}

To illustrate the influence of the chosen model chemical system, we solve the inverse problem for the case of the Earth
as in section 3.2, but using the simpler FMS system.
Sampled model parameter pdfs are shown in Figure \ref{fms_earth} and 
relative to NCFMAS (Figure \ref{figure_earth}), we observe that with the FMS system
mantle composition and core radius are less well-predicted, although posterior model parameter variability is,
as expected, smaller (fewer chemical components).
Also, future use of stellar determinations of Ca, Na, and Al abundances support the use of the NCFMAS system.

\section{Shannon entropy measure}
\label{shannon}

The Shannon entropy measure of a discrete variable $X$ of possible states $\{x_1,x_2,..,x_{N}\}$, with $N$ being the number of bins, can be written as:

$$
H = - \sum_{i=1}^{N} p({x_i}) * {\rm log_2} p({x_i})
$$
where the $x_{i}$ states represent the distribution histogram bins in which the probability of occurrence is given by
$$
p({x_i}) = \frac{\text{ \# of realizations in the bin}}{\text{total \# of realizations}}
$$

The Shannon entropy $H$ has a maximum $H_{\rm max} = { \rm log_2}(N)$ when all $N$ states are equally likely (e.g., a uniform distribution) 
and a minimum
$H_{\rm min} = 0$ when all realizations $x_i$ fall into a single histogram bin. Clearly, $H$ is bin-size dependent. Here, the bin size used is defined such that the prior range of each model parameter is divided in $\approx$ 20 different bins.

\bibliographystyle{aa} 
\bibliography{Mybibfile}

\end{document}